\documentclass{article}
\usepackage{latexsym, amsmath}

\everydisplay{\rm }
\everymath{\rm}
\begin{document}
\begin{center}
\LARGE Spin network coherent states for planar gravitational waves
%the \\[.25in] leaves .25 in of blank space then starts a new line
\\ [.25in] \large Donald E. Neville \footnote{\large Electronic
address: dneville@temple.edu}
\\Department of Physics
\\Temple University
\\Philadelphia 19122, Pa. \\ [.25in]
April 29, 2013 \\ [.25in]
\end{center}

%%%%%%Ashtekar Notations %%%%%%%%%%%
\newcommand{\E}[2]{\mbox{$\tilde{{\rm E}} ^{#1}_{#2}$}}
%Enter \E{A}{b} to get E super A sub b
\newcommand{\A}[2]{\mbox{${\rm A}^{#1}_{#2}$}}
\newcommand{\K}[2]{\mbox{${\rm K}^{#1}_{#2}$}}
\newcommand{\Np}{\mbox{${\rm N}'$} }
\newcommand{\Etwo}{\mbox{$^{(2)}\!\tilde{\rm E} $} }
\newcommand{\Etld }{\mbox{$\tilde{\rm E}  $}\ }
\newcommand{\Vtwosq}{\mbox{$(^{(2)}\!{\rm V})^2 $} }
\newcommand{\Vtwo}{\mbox{$^{(2)}\!{\rm V} $}\ }
\newcommand{\Vthree}{\mbox{$^{(3)}\!{\rm V} $} }

%%%%%%%Tex Newcommands%%%%%%%%%
\newcommand{\bea}{\begin{eqnarray}}
\newcommand{\eea}{\end{eqnarray}}
\newcommand{\be}{\begin{equation}}
\newcommand{\ee}{\end{equation}}
\newcommand{\nn}{\nonumber \\}
\newcommand{\rta}{\mbox{$\rightarrow$}}
\newcommand{\rla}{\mbox{$\leftrightarrow$}}
\newcommand{\eq}[1]{equation~(\ref{#1})}
\newcommand{\Eq}[1]{Equation~(\ref{#1})}
\newcommand{\eqs}[2]{equations~(\ref{#1}) and (\ref{#2})}
\newcommand{\Eqs}[2]{Equations~(\ref{#1}) and (\ref{#2})}
%call this using two sets of brackets: \eqs{first}{second}
\newcommand{\bra}[1]{\langle #1 \mid}
\newcommand{\ket}[1]{\mid #1 \rangle}
\newcommand{\braket}[2]{\langle #1 \mid #2 \rangle}
\large
\begin{center}
{\bf Abstract}
\end{center}
    This paper constructs
coherent states for spin networks with
planar symmetry. After gauge-fixing, the full SU(2) symmetry is
broken to U(1); but one cannot simply use the U(1) limit of
SU(2) coherent states, because the planar states exhibit an
unexpected O(3) symmetry arising from the closed loop character of the
transverse directions.  The coherent states constructed in
this paper obey this symmetry.  They are superpositions of
holonomies which obey the U(1) gauge symmetry only on
average: some holonomies in the superposition violate U(1)
symmetry, although the U(1) quantum numbers of the state are peaked at
values which obey the symmetry. Operators acting on coherent
states give back a c-number times the original state,
plus small correction states, which make the coherent state
an approximate rather than exact eigenstate of the operator.
In a follow-on paper these small correction states are
used to calculate small corrections to eigenvalues
of the volume operator.
\\[.125in]
PACS categories: 04.60, 04.30

\clearpage

\section{Introduction}\label{SecIntroduction}

    This paper is based on the canonical, spin network
formulation of quantum gravity.  This approach formulates gravity using
densitized triads and a connection,
rather than a metric \cite{Ash,Sen}.  The connection is real
\cite{Barbero}, and exponentiated to form a
a holonomy.

    To construct the full theory, one must choose a Euclidean
Hamiltonian and consider commutators between holonomy,
volume, and Euclidean Hamiltonian \cite{QSD}.  Since commutators
with the Hamiltonian are also essentially [volume, holonomy]
commutators, the essential commutator is between the holonomy
and the triads making up the volume operator.  Not only the Hamiltonian,
but also the states are constructed from holonomies;
therefore it is desirable to make the action of triads on holonomies
as simple as possible; this is one goal of the present paper.

    The holonomy has its support, not on a
continuous manifold, but rather  on a one-dimensional network of
edges meeting at vertices (spin network) \cite{JacobSmo,RovSmo}.
At present there is no empirical confirmation for the existence
of a microscopic spin network structure,
although in the future it may be possible to detect Planck-scale
modifications to elementary particle decay rates, or modifications
to light propagation over cosmological distances \cite{GambPull, AMTU,
LJM,Bluhm}.  For now, the approach must be checked by
undertaking calculations which confirm the internal consistency of the
formulation, or confirm consistency with established principles.  For example,
area and volume operators for the
theory possess a discrete spectrum \cite{discrAV}, in qualitative agreement with the
conclusions from thought experiments that lengths smaller than a Planck
length are not measurable \cite{minL}.  Cosmological calculations, extended back to
the big bang, yield a finite result \cite{Bojcosm}.  Black hole
calculations predict area $\propto$ entropy and yield a formula for the
entropy \cite{SeqA1,SeqA2}.

    One topic which requires further investigation is the classical
limit of the spin network
approach.  The systems studied so far (black holes, homogeneous
cosmologies) have reasonable classical limits,
but they are so highly symmetric that they cannot propagate
gravitational waves.

    In order for the system to propagate waves, the
spin network Hamiltonian must be nonlocal:
a single term in the Hamiltonian must be able to change the spin network
at two or more neighboring vertices \cite{SmolLR}.
Once the Hamiltonian is made nonlocal, it is no longer obvious that
the constraints have the correct commutation relations, in the limit of
fields varying slowly over many spin network vertices \cite{NevLR}.
I.~e., it is not
obvious that the theory possesses general coordinate invariance in
the classical limit.

    Several approaches use the path integral,
or "spin foam" approach to study the classical limit,
rather than the canonical, spin network approach.
Aleschi and Rovelli calculate the gravitational Green's function, then
check for a correct long-range behavior \cite{Ggrav}.  This approach puts limits on
allowable spin foam vertices.  Ambjorn, Jurkiewicz, and Loll put the system
in a heat bath and solve numerically for the ground state, to see
if the geometry is reasonable \cite{HeatBath}. This work puts limits on the topology of
admissible paths, and favors spin networks which are causal.  Markopoulou,
and also Oeckl, have studied the application of the
renormalization group to gravity theory\cite{RG1,RG2}.

    Thiemann and Winkler develop another approach to the classical
limit: within the canonical approach, construct coherent states which can be used
to study the theory approximately in the classical limit
\cite{GCSI,GCSII, GCSIII}.  This is the approach
followed in the present paper, which constructs coherent states for a space
possessing two commuting
spacelike Killing vectors which may be written $\partial /\partial x$ and
$\partial /\partial y$.  This planar symmetry
is the simplest which allows propagation of gravitational waves,
and therefore requires a nonlocal Hamiltonian.

    For this case one cannot
use the coherent states constructed by Thiemann and Winkler. In one sense
these states are too general.  They
possess the full SU(2) local gauge invariance, whereas planar symmetry
allows the gauge to be fixed, until only U(1) gauge rotations around the Z axis
survive \cite{HusSmo}.  (Lower case coordinates x,y,z,t refer to coordinates on the global
manifold; upper case coordinates X,Y,Z,T refer to coordinates in local
Lorentz frames.) In another sense, the Thiemann Winkler states are not
quite general enough, because the planar states
must possess an O(3) symmetry which is not a limit of the SU(2) gauge
symmetry \cite{VolI,VolII}.

    Section \ref{SecHilbert}
describes a suitable Hilbert space for plane waves, and discusses the need for
the O(3) symmetry.  This section
is a summary of an earlier paper by the author \cite{VolI}, and is included
to make the present paper reasonably self-contained.

    Section \ref{SecCoh} constructs the coherent states.
The approach used in that section is rather intuitive.  However,
in later sections I show explicitly that the states have the
properties expected of coherent states.  They are approximate eigenvectors
\[
    O\ket{\mbox{coherent state}} = <O>\ket{\mbox{coherent state}} + SC,
\]
where the operator O is either a densitized triad \E{I}{i} or a
spin one-half holonomy.  The leading contribution
to the matrix element is a c-number function $<O>$; and SC is a small
correction term.  Section \ref{SecEtld} derives $<O>$ for the densitized
triad; section \ref{SecHolonomy} derives it for the holonomy.
Quantitative estimates of the SC terms are worked out
in the appendices to this paper.  The volume
operator is a key operator in the spin network approach.  Section
\ref{SecVhcomm} considers the commutator [ holonomy, volume ].

    The present paper is primarily an explication of the
properties of coherent states, independent of any choice of Hamiltonian;
however, I have emphasized certain properties of coherent
states, because these are the properties I expect will be relevant
later, when calculating the Hamiltonian.  In particular, I have assumed
that any future Hamiltonian will be constructed from spin 1/2 holonomies,
those in the fundamental representation of SU(2).
Although  spin 1/2 is the simplest choice, Gaul and Rovelli
have investigated Hamiltonians involving
higher spin \cite{Gaul}.  They find there is no problem
in principle with using the higher holonomies.  Also, there is no
problem extending the present formalism to higher holonomies, if they
are needed.

    The coherent states constructed in this paper are not exact eigenstates of
the residual U(1).  Exact eigenstates could be
obtained by angle-averaging the coherent states which will be constructed in section \ref{SecCoh}.
But these coherent states are already quite complicated, even before an angle average.
It seems simplest to test for general covariance and
gravitational wave propagation, using states
which obey U(1) only on average, initially; the
calculation can be refined at a later point.  For studies of angle-averaged
states which obey SU(2) exactly, see
reference \cite{exactSU2}.

\section{The planar Hilbert space} \label{SecHilbert}

    In reference \cite{VolI} I proposed a spin network for
the planar case, and constructed a kinematic basis.
For completeness I review the highlights of that construction
here.

    Call the direction of propagation the z direction.
The spin network in the z direction has the expected topology, a
series of vertices connected by edges in the z direction.
Holonomies on the z axis look like holonomies in the full
theory.  Each holonomy is integrated from one vertex to the next.

    Each vertex also has an infinite number of vertices stretching in
the x and y directions, but because of the symmetry, the holonomy
stretching from vertex n to vertex n+1 is identical to the holonomy
stretching from n-1 to n.  Rather than an infinite number of vertices,
one can bend the n to n+1 holonomy
around in a circle and associate both ends of this holonomy with the
same vertex n.  I.e. the x and y edges may be
given the topology of a circle.

    I will
work in a connection representation for the wave function.  The
wave function at each vertex is a product of four holonomies:
the two x and y holonomies on the circular edges, plus one incoming z
holonomy and one outgoing z holonomy.

    These holonomies may be simplified by gauge-fixing
the \Etld and connection
fields \cite{HusSmo}. The off-diagonal elements \E{a}{Z} and
\E{z}{A}, with a =
x,y and A = X,Y, can be gauged to zero; similarly, \A{Z}{a} and
\A{A}{z} may be set to zero.

    This means that the holonomies along
the longitudinal z direction are quite simple, involving only the
\A{Z}{z} and the rotation generator $S_Z$ for U(1) rotations around Z.
Coherent states for the case of U(1) symmetry are well understood;
see for example Thiemann and Winkler \cite{GCSIII}.
The basis holonomies along z are

\begin{equation}
    h_z = \exp \,[\,i \int M_z \,A^Z_z \, dz \,],
    \label{defhz}
\end{equation}
where $M_z$, an eigenvalue of the diagonal generator $S_z$, is integer
or half-integer.

    Now consider the x and y holonomies.  Since \A{Z}{a} and
\A{A}{z} have been set to zero, these involve generators $S_X,S_Y$ and are
rotations in the X,Y plane.
Each transverse holonomy,
\[
 h^{(1/2)}  = \exp[ \,i \,\hat{m} \cdot \vec{\sigma}\,\theta/2 \,],
\]
therefore has an axis of rotation
of the general form

\begin{equation}
        \hat{m} = (\cos \phi, \sin \phi,0),
        \label{defm}
 \end{equation}
for some angle $\phi$. (More precisely, there is one holonomy for
each transverse direction x,y; and one $\phi$ for each transverse
direction, $\phi_x$ and $\phi_y$.  Since the two directions are
treated equally, I will discuss only the x holonomies, and will
suppress the subscript x for now.)

    The spin 1/2
holonomy $h^{(1/2)}$ has matrix elements
\begin{equation}
h^{(1/2)}
    =\left[\begin{array}{cc}
        \cos (\theta/2) & i \exp(-i\phi)\sin (\theta/2) \\
        i\exp(+i\phi)\sin (\theta/2)& \cos (\theta/2)
        \end{array}
        \right] .
    \label{defh}
\end{equation}
The usual Euler angle decomposition is
\begin{align}
    h^{(1/2)}& =  \exp[-i\sigma_Z(\phi-\pi/2)/2 \,] \,\exp[ \,i\sigma_Y
    \theta/2 \,] \nn
    &\cdot \exp[+i\sigma_Z(\phi-\pi/2)/2 \,] \nonumber \\
   & =  h^{(1/2)}(-\phi+\pi/2,\theta,\phi-\pi/2) .
\label{EulerDecomp}
\end{align}

    The natural basis for the transverse
Hilbert space  might seem to be the generalization of $h^{(1/2)}$
from 1/2 to general j, the set of rotation matrices

\[
    h^{(j)}(-\phi+\pi/2,\theta,\phi-\pi/2),
\]
where j is the highest weight obtained by multiplying together 2j
$h^{(1/2)}$ matrices. However, this basis is not convenient
because it has complicated behavior under the action of the \Etld and
the volume operator.  For example, for j = 1/2,  \Etld (acts as a
functional derivative with respect to A and)
produces an anticommutator.

\begin{eqnarray}
\E{x}{A} \,h^{(1/2)} &=& \E{x}{A} \exp[ \,i\int \A{B}{x} S_B \, dx \,]\nn
    &=& (1/2)\, \gamma \kappa \, [ \,\sigma_A/2, \: h^{(1/2)} \,]_+,
\label{actionE}
\end{eqnarray}
This anticommutator shuffles the matrix elements of $h^{(1/2)}$ in a
complicated way.  The anticommutator arises because the
transverse holonomy is supported by an edge with the
topology of a loop: the holonomy both begins and ends at the same
vertex.  Therefore \Etld "grasps" the holonomy on both
sides.

    In \eq{actionE} $\kappa = 8 \pi G; \gamma$ = Immirsi parameter; and the 1/2
comes about because the \Etld grasps the $\int A\cdot S$ argument of
the holonomy at endpoints , resulting in half a delta function.
The delta functions are always canceled by the area and line
integrals associated with \E{x}{A} and \A{B}{x}.  I have suppressed
the area integration associated with each \Etld.

    Fortunately, the \Etld reshuffle the elements of h in a relatively
simple way.  Introduce the operators \E{x}{\pm}, where as usual

\begin{equation}
    f_{\pm} := (f_x \pm i f_y)/\sqrt{2}.
    \label{defpm}
\end{equation}
The operators \E{x}{\pm} reshuffle the components of h in the
same way that the familiar angular momentum operators $L_{\pm} $
reshuffle the L=1 Legendre polynomials $Y^M_1$.  For example,
write out the action of
the anticommutator in \eq{actionE}, for index A = +.
\begin{equation}
    [ \,\sigma_+/2, h^{(1/2)} \,]_+
        = \sqrt{1/2}\left[\begin{array}{cc}
        i\exp(-i\phi)(\sin \theta/2) & 2 \cos )\theta/2) \\
        0               & i \exp(+i\phi)\sin (\theta/2)
        \end{array}
        \right]
    \label{E+h}
    \end{equation}
Compare this matrix to the original matrix, \eq{defh}:
\E{x}{+} has reshuffled the matrix elements as follows
\begin{eqnarray}
    (i/\sqrt{2}) \exp(-i\phi)\sin \theta/2 &\rightarrow&  \cos \theta/2 \nn
       \cos \theta/2 &\rightarrow & (i\sqrt{2}) \exp(+i\phi)\sin \theta/2 \nn
      (i/\sqrt{2}) \exp(+i\phi)\sin \theta/2  &\rightarrow& 0.
\label{E+Shuffle}
\end{eqnarray}
This is isomorphic to the action of the operator $L_+ $ on the
L = 1 Legendre polynomials.  The isomorphism is
\begin{eqnarray}
    L_{\pm} &\leftrightarrow& 2 \,\E{x}{\pm}/\gamma \kappa; \nn
    L_0 &\leftrightarrow& 2 \,\E{x}{0}/\gamma\kappa; \nn
    Y^{\pm}_1(\theta,\phi) &\leftrightarrow& \,Y^{\pm}_1(\theta/2,\phi - \pi/2)\nn
                            &=& \mp N \sin(\theta/2)\exp[ \,\pm (i\phi - i \pi/2) \,]/\sqrt{2};\nn
    Y^{0}_1(\theta,\phi) &\leftrightarrow&  Y^{0}_1(\theta/2,\phi - \pi/2)\nn
                        &=& N \cos (\theta/2).
\label{isomorphism}
\end{eqnarray}
Because of the half angles, the new Y's are orthonormalized
using
\[
    \int_0^{2 \pi} \,\int_0^{2 \pi} \,Y^* Y \,\sin(\theta/2) \,d(\theta/2) \,d \phi.
\]
After a change of variable $\rho = \theta/2$, this dot product is the
usual one.  N is the usual normalization $\sqrt{4\pi/3}$.

     Because the $Y^M_1(\theta/2,\phi - \pi/2)$
transform more simply than matrix elements of h under the action of \Etld, one
obtains a more convenient basis by using O(3) Clebsch-Gordan coefficients to
join the Y's, rather than SU(2) coefficients to join elements of h.  The
resultant basis is just the set of spherical harmonics $Y^M_L (\theta/2,\phi - \pi/2)$
for O(3).

    The operator \E{x}{0} is not
the triad \E{x}{Z}, which has been gauged to zero.
\E{x}{0} is constructed to complete the trio
of generators and act on states in a manner isomorphic to $L_0$.

\begin{equation}
    \E{x}{0} = [ \,h, \: \sigma_3/2 \,]_-
\label{defEsubZero}
\end{equation}

    Because the three independent elements of $h^{1/2}$ can be expressed in
terms of the $Y^M_1(\theta/2,\phi - \pi/2)$, the Y's are as complete
a set as the elements of $h^{1/2}$.  The relation between h and the $Y^M_1$
is

\begin{equation}
    N\mathbf{h}^{(1/2)} = \mathbf{1}Y^0_1 + i Y^+_1 \boldsymbol{S_-} +i Y^-_1 \boldsymbol{S_+} ,
        \label{heqY}
\end{equation}
where boldface denotes a 2x2 matrix. This expansion of ${h}^{(1/2)}$
contains only three matrices and no $\sigma_3$, which explains
why h contains only three independent elements, corresponding to the
three components $Y^M_1$.

    One can take into account the y edges as well as the
x edges, by constructing two bases, $Y^{Mx}_{Lx}$ and
$Y^{My}_{Ly}$ for holonomies along the x and y directions
respectively. These harmonics transform simply under the action
of the \Etld:
 \be
    (\gamma\kappa/2)^{-1}\E{x}{\pm} \,Y^M_{L} =
            \Sigma_{N} Y_{L N} \,\bra{L,N} S_{\pm}\ket{ L,M},
                \label{Ytransf}
 \ee
where $Y_{L M} = Y_{L M}(\theta /2,\phi -\pi /2)$.  The
unconventional half-angle reminds us of the origin of these
objects in a holonomy $h^{1/2}$ depending on half-angles.

    The Y's are known to be proportional to matrix elements
of rotations,

\begin{equation}
    \sqrt{4\pi/(2L+1)} \,Y_L^M =
            D^{(L)}_{0M}(-\phi +\pi /2, \theta /2,\phi -\pi/2).
                \label{YeqD}
\end{equation}
Therefore \eq{Ytransf} is also correct if D's are substituted for
Y's. I prefer D's to Y's in what follows, because use of D's (will
require awkward factors of $\sqrt{4\pi/(2 \,L+1)}$ in initial
formulas, but) ultimately will result in fewer factors
$\sqrt{4\pi/(2 \,L+1)}$.

    The planar calculation involves three groups: SU(2), because
the holonomies in the Hamiltonian are rotation matrices in SU(2); O(3),
because the action of the grasps generates an O(3) group; and
U(1), because the usual SU(2) gauge invariance is broken to U(1)
by the gauge fixing.  It is worth taking a minute to contemplate
when to use which group.

    Presumably the  Hamiltonian will be constructed using
holonomies $h^{(1/2)}$ in the fundamental representation of SU(2).
The Euler decomposition of $h^{(1/2)}$, \eq{defh}, shows $h^{(1/2)}$
as depending on the full angle $\theta$, but of course the actual
matrix elements of $h^{(1/2)}$ contain half-angles $\sin(\theta /2),\,
\cos (\theta /2)$.  When these matrix elements are rearranged to form
a representation $D^{(1)}(h)$ of O(3), the matrix elements of  $D^{(1)}(h)$
(and $D^{(L)}(h)$, \eq{YeqD}) inherit this dependence on half-angles:
the Euler expression for the D's involves a half angle.
I will use the notations h for the SU(2) matrix (as, $h^{(1/2)}$)
and D(h) for the O(3) matrices.

    Now consider the relation between O(3) and the residual
gauge invariance U(1).  Most of the O(3) rotations have nothing
to do with SU(2) gauge invariance.
The exceptions are rotations around the axis of propagation Z.
These rotations are just the U(1) gauge rotations.  Proof:
because of the gauge fixing,
the axis of rotation $\hat{m}$ for the matrix $h^{(1/2)}$,  \eq{defm} lies in the XY
plane, and must remain in that plane.
The SU(2) gauge rotations of the holonomies are reduced to
U(1) rotations around the Z axis.  The result of a Z rotation
of the connections follows from the surviving Z component of Gauss.

\begin{eqnarray*}
    \A{X}{x}' &=& \A{X}{x} \cos \delta - \A{Y}{x} \sin \delta;\\
    \A{Y}{x}' &=& \A{Y}{x} \cos \delta + \A{X}{x} \sin \delta:\\
    \A{Z}{z}' &=& \A{Z}{z} - \partial_z \delta.
\end{eqnarray*}
This induces a rotation of the holonomies,

\begin{eqnarray*}
    h' &=& \exp[-i\sigma_z \delta /2 \,] \,h \,\exp[ \,+i\sigma_z \delta /2 \,];\\
    h'_z &=&\exp [ \,i \int_1^2 A'_z \,dz \,M_z \,] \\
         &=& \exp[ \,-i M_z \,\delta(2) \,] \,h_z \,\exp[+i M_z \,\delta(1) \,].
\end{eqnarray*}
The first line, above, implies that the argument $\phi$ of h changes
to $\phi + \delta$.

\begin{equation*}
     h'^{(1/2)}_{mn} = h^{(1/2)}_{mn}
                            \exp[ \,i\, (n-m)\,\delta \,].
\end{equation*}
This in turn changes the components of the $Y^M_L$.
\begin{equation*}
     D'^{(L)}_{0M} = D^{(L)}_{0M} \,\exp[ \,i\, M \,\delta \,].
\end{equation*}
This is identical to the result of an O(3) rotation
around the Z axis. $\Box$

    It is straightforward to work out the consequences of
U(1) gauge invariance for the simplest basis, where the x and y
holonomies have definite M, $D^{(La)}_{0Ma}$, a = x,y, and the incoming and
outgoing z holonomies, equation \eq{defhz}, have definite $M_{zf}$
and $M_{zi}$.  The total change in phase at each vertex is

\[
    \exp[ \,i \delta  (M_x  + M_y + M_{zf} - M_{zi}) \,].
\]

U(1) invariance therefore demands that the quantity
\[
        M_x  + M_y + M_{Zf} - M_{Zi}
\]
vanish, for each vertex.

    The transverse coherent states constructed here will not have
unique values for $M_x$
and $M_y$.  These states will be superpositions of  $D^{(La)}_{0Ma}$
matrices; and the superpositions will contain a range of values $M_a$.
(Similarly, coherent states in the longitudinal direction will not have
definite $M_z$.)  The superpositions are sharply peaked at central
values of the M's, however,
so that M-values which violate U(1) are suppressed.

\section{Coherent states}\label{SecCoh}

    From the previous
section, a basis for the Hilbert space can be constructed
from O(3) spherical harmonics
\[
    D^{(1)}_{0M}(-\phi +\pi /2, \theta /2,\phi -\pi/2) := \,
            D^{(1)}(h)_{0M},
\]
A set of coherent states should therefore be a sum over spherical
harmonics,

\begin{equation}
    \ket{u,\vec{p}} = \sum_{L,M} D^{(L)}_{0M}(h) \,c(L,M;u,\vec{p}),
\label{cohprops}
\end{equation}
for some coefficients c.
The parameters (u,$\vec{p}$) label the
coherent states.  u is a 2x2 matrix specifying the peak
value of h, and $\vec{p}$
determines the peak value of \E{x}{A}.

    There are two kinds of SU(2) coherent states available in the
literature.  The first type has no sum over L in \eq{cohprops},
only a sum over M \cite{Radcliffe,Perolomov}.  These coherent
states are too simple for
present purposes.  If L is fixed, the \Etld operators can be made
classical, but not the holonomies.  Since L is conjugate to angle,
one must superimpose many L's
to get sharply peaked angular values for $(\theta,\phi)$.

    The second kind of coherent state was suggested by Hall
\cite{Hall} and
elaborated for quantum loop gravity by Thiemann and Winkler
\cite{GCSI,GCSII,GCSIII}.
Their results were derived for the general case, full local SU(2)
symmetry.  For the planar case, with its O(3) symmetry,
simply taking a U(1) limit of the general case does not work.
One must start from a superposition of representations of
the O(3) symmetry, as in \eq{cohprops}

    The Thiemann Winkler coherent states may be
understood intuitively as generalizations of the minimal
uncertainty states for the free particle. One applies a
certain recipe to construct the free particle states, and the same
recipe works in the general SU(2) case. Once this intuitive
approach is understood, it is straightforward to use the same
approach to generate a set of candidate coherent states for
the planar O(3) case.

    I review the recipe for constructing a
coherent state for the free particle. Start from a wave function
which is a delta function.

\[
    \delta (x-x_0) = \int exp[ \,ik(x-x_0) \,] \, dk/2\pi.
\]
This wave function is certainly strongly peaked, but it is not
normalizable. Also, it is peaked in position, but it should be
peaked in both momentum and position.  To make the packet
normalizable, insert a Gaussian operator $\exp[ \,-p^2/(2\sigma^2) \,]$.
(Choosing the Gaussian form is a "cheat", because we know the
answer; but for future reference note that all the eigenvalues
$k^2$ of the  operator $p^2$ must be
positive, so that the Gaussian damps for all k.) To produce a
peak in momentum, complexify the peak position: $x_0 \rightarrow
x_0 + i p_0 /\sigma ^2$.  With these changes, the packet becomes

\begin{multline}
    N\int  \exp [-p^2/(2\sigma^2) \,] \exp[ \,i k(x-x_0) + k p_0/\sigma^2 \,] \, dk/2\pi\\
        =  N\int \exp [ \,-k^2/(2\sigma^2) + i k(x-x_0) + k p_0/\sigma^2 \,]\, dk/2\pi\\
        =(N \exp [ \,p_0^2/(2 \sigma^2) \,]/\sqrt{2\pi}) \\
        \cdot \exp[ \, ip_0(x-x_0) -(x-x_0)^2 \sigma^2/2 \,].
        \label{freecoh}
\end{multline}
The last line follows after completing the square on the
exponential, and exhibits the characteristic coherent state form.

    There is not just one state, but a family of coherent
states, characterized by the parameter $\sigma$.  The shape of the
wave function is highly sensitive to $\sigma$; but the peak values
$x_0, p_0$ are independent of $\sigma$, as is the minimal
uncertainty relation $\Delta x \Delta p = \hbar /2$.  The
coherent states constructed below contain a parameter t which is
analogous to $\sigma$.

    Now apply the above recipe to the planar case.  The conjugate
variables x and p are replaced by a pair of conjugate variables:
angles $(\theta,\phi)$ and angular momenta (L,M). The complete
set of momentum eigenfunctions is replaced by a complete
set of spherical harmonics.

    To construct a delta function in angles,
\[
    \delta (\theta/2 - \alpha/2)\delta(\phi - \beta)/\sin (\alpha/2)
\]
I introduce spherical harmonics $D^{(L)}(u)$ depending on angles
$(\alpha,\beta)$ in the same way that the $D^{(L)}(h)$ depend on
$(\theta,\phi)$.

\bea
      D^{(L)}(u)_{0M} &=& D^{(L)}(-\beta+\pi/2,\alpha/2,\beta-\pi/2)_{0M}\nn
               &=&\sqrt{4\pi/(2L+1)} \,Y_{L M}(\alpha/2,\beta -\pi /2).
            \label{defDu}
\eea
Compare \eq{defDu} to \eq{YeqD}.  I can now write the
delta function in angle as a sum over spherical harmonics.

\begin{multline}
    \delta (\theta/2 - \alpha/2)\delta(\phi - \beta)/\sin (\alpha/2) \\ =
        \sum_{L,M}((2 \,L+1)/4\pi) \,D^{(L)}(h)_{0M} \,D^{(L)}(u)_{0M}^* .
        \label{deltaangle}
\end{multline}
As discussed in the last section, it is more convenient to use D
matrices rather than $Y_{LM}$'s, but the reader who wishes to exhibit
the latter can use \eqs{YeqD}{defDu} to replace D's by Y's in
\eq{deltaangle}. The sum then takes on a form which may be more
familiar, just $\sum Y Y^*$.  The momentum eigenvalue k in the free
particle example corresponds to angular momentum eigenvalue L in the planar case.

    Continue with the recipe: dampen the sum using a Gaussian $\exp [-t
L(L+1)/2 \,]$.  Complexify by extending the angles in u to complex
values, replacing u by a matrix g in the complex extension
of O(3).  (For the free particle, x becomes complex; here, the
angles become complex.)  The coherent state has the general form

\begin{align}
    \ket{ u,\vec{p}} = & N \sum_{L,M}((2 \,L+1)/4\pi)
                    \exp[-t L(L+1)/2 \,]  \nn
                    & \cdot D^{(L)}(h)_{0M} \,D^{(L)}(g)_{0M}^*
                    \nn
                    = & \, N
                    \sum_{L,M}\cdots D^{(L)}(h)_{0M} \,D^{(L)}(g\dag)_{M0}.
\label{defcoh}
\end{align}

    A vector $\vec{p}$ is needed to characterize the matrix g, as
follows. Every matrix in SL(2,C), the complex extension of
SU(2), can be decomposed into a product
of a Hermitean matrix H times a unitary matrix u ("polar
decomposition"; see for example \cite{Halltext}).  E.~g. for
the fundamental representation,
\[
    g = H u = \exp [\,\vec{\sigma}\cdot \vec{p}/2 \,] \,u.
\]
It follows
that every matrix in O(3) also has a polar decomposition, obtained
by restricting the representations of SU(2) to representations
with integer spin.

\bea
    g^{(L)} &=& \exp [ \,\vec{S}\cdot \vec{p} \,]  \,u^{(L)} \nn
        & := & H^{(L)} \,u^{(L)}
        \label{geqHu}
\eea

    At this point I must choose six input parameters: the three
Euler angles which determine the unitary matrix u and the three
components of $\vec{p}$, which define the complex extension.
By analogy with the free particle case, if u determines the
peak angles, then the complex extension $\vec{p} \,$ determines
peak values of the canonically conjugate variable, the
angular momentum.

    In this paper I use the following
choices for these parameters.  Restrict $\vec{p} \,$ to be an arbitrary
vector in the XY plane.

\be
    p_3 = 0.
    \label{p3eq0}
\ee
Restrict the unitary matrix u to be an arbitrary rotation with
axis of rotation in the XY plane.  The Euler decomposition of
$u^{(L)}$ is

\bea
    u^{(L)} &=& u(-\beta + \pi/2,\alpha,\beta - \pi/2)\nn
             &:=& \exp[ \,i \,\hat{n} \cdot \vec{S}\,\alpha /2 \,];\nn
    \hat{n} &=& (\cos \beta, \,\sin \beta,\,0).
           \label{defu}
\eea
(Compare this definition of u to the corresponding definition
of the matrix h, given at \eqs{defm}{defh}.  Both u and h have their
axis of rotation in the XY plane.)

    These choices are certainly plausible.  In the limit
of no complexity (in the limit
$\vec{p}\rightarrow 0$) the coherent state will reduce to the
original delta function and u will become the peak value of h.
Since h has its axis of rotation in the XY plane, the peak value
should also be a matrix with axis in the XY plane.

    Also, the generator $\vec{S} \cdot \vec{p} $ for H is the
complexification of the generator $\vec{S} \cdot \hat{n} $ for
u.  Since $\hat{n}$ lies in the XY plane, the same should
be true for $\vec{p}$.  In short, since h is a very special matrix
(its axis lies in the XY plane) the peak matrix u and its
complexification H should also be special.

    I have considered more general choices for u and $\vec{p}$,
\[
    (u, p) \rta (\tilde{u},\tilde{p}),
\]
where $\tilde{p}$ and the axis of $\tilde{u}$  need not be in
the XY plane. However, in later sections of this paper I
calculate the expectation values
$<\E{x}{A}>$ and \mbox{ $<D^{(1)}(h)_{0A}>$}, for
$p_3 = 0$, and axis of u in the XY plane, \eqs{p3eq0}{defu},
and find that these simpler choices already give coherent
states with the required peaked behavior.
Use of the more general values seems to add
nothing but complexity.

   Since $\vec{p} \,$ now lies in the XY plane,
$\vec{p} \,$ may be parameterized using a magnitude, p, plus an
angle $\mu$.  It is convenient for later calculations to take
 $\mu$  to be the angle between $\vec{p} \,$ and $\hat{n}$, the axis
of u.

\be
    \vec{p} = p \,[\cos(\beta +\mu), \,\sin(\beta +\mu), \,0].
        \label{defmu}
\ee
From \eq{defu}, $\beta$ is the angle between $\hat{n}$ and the X
axis.

\section{Overview}\label{SecGenForm}

    The next two sections verify the properties given in \eq{cohprops}. The
general structure of the
calculations is relatively simple, but there are many
details which can obscure matters.  It may be useful to summarize
that structure here, to avoid losing the forest for the trees.

    The peak values of both triad and holonomy,
\[
    <\E{x}{A}> \:\mbox{and} \: <D^{(1)}(h)_{0A}>,
\]
are both vectors in SU(2)
index space (A = X,Y,Z).  It is desirable to construct an
orthonormal triad of unit vectors with
physical significance; then express
the two peak values in terms of these unit vectors.

    An obvious choice for one unit vector is
the peak value of $D^{(1)}(h)_{0A}$.
From the way the
delta function in angle was constructed at \eq{deltaangle},
the peak value of holonomy $D^{(1)}(h)_{0A}$ will turn out to be
$D^{(1)}(u)_{0A}$:

\[
    <D^{(1)}(h)_{0A}> = D^{(1)}(u)_{0A}.
\]
Every row of an orthogonal matrix is a unit
vector, and in particular
the three matrix elements $D^{(1)}(u)_{0A}$ form
a unit vector.  I introduce a notation which
emphasizes the unit character of $D^{(1)}(u)_{0A}$:
\begin{equation}
    D^{(1)}(u)_{0A} := \hat{D}_A.
\label{defDhat}
\end{equation}

    The vector $\hat{n}$, the axis of rotation
for $D^{(1)}(u)$, is another choice for a unit
vector.  It is is orthogonal to $\hat{D}$.  To derive
the orthogonality,  write $\hat{D}$ as the rotation
of a unit vector initially along the Z axis.
\begin{equation}
    \hat{D}_A = (0,0,1)_B D^{(1)}(u)_{BA}.
\label{DeqHolonomy}
\end{equation}
The initial unit vector (0,0,1) = $\hat{Z}$ is orthogonal to
the axis of  rotation $\hat{n}$, since the latter lies in the
XY plane. It follows that the rotated unit vector,
$\hat{D}$, remains orthogonal to $\hat{n}$, since
orthogonal rotations preserve angles.  $\Box$
Given the two orthogonal basis vectors, ($\hat{n}, \,\hat{D}$), one
can construct a third from $\hat{n} \times\hat{D}$.

    Action of  the operators on the coherent state does not
lead to the three unit vectors immediately, but does give
rise to a factor $D^{1}(u)_{BA}$.  The row B = 0 yields
$\hat{D}$.  The B = $\pm 1$ rows are orthogonal to the B = 0
row, and appendix \ref{AppD1u}  expresses them in terms of $\hat{n}$ and
$\hat{n} \times\hat{D}$.

    The $D^{1}(u)_{BA}$ are multiplied by  factors of
$D^{(L)}(H)_{M0} \,\exp[-t L(L+1)/2 \,]$.  From appendix
\ref{AppDH} this factor has a Gaussian form for large $e^p \gg 1$.
\begin{multline}
    \exp(-t L(L+1)/2 \,) \,D(H)^{(L)}_{MN} \nn
    \cong \exp[-t((L+1/2)-p/t)^2 /2 \,]
             \nn
     \cdot \exp [-M^2 /2(L+1/2) -N^2 /2(L+1/2) \,]
                [1/\sqrt{\pi (L+1/2)}]\nn
     \cdot \exp [ p^2/(2t) -p/2](\exp[-i(\beta + \mu)])^{M-N}.
            \label{DHforp3eq0}
\end{multline}
This result suggests power-series expanding the
complex expressions around the mean values supplied by the
Gaussian, L + 1/2 = p/t and M = 0.  The lowest powers supply
the peak values; higher powers can be used to
estimate the size of the small correction (SC) terms.

%    The other vector used to
%characterize the coherent state, the vector $\hat{p}$,
%is less convenient to use (as basis) than $\hat{n}$.
%$\hat{D}$ has components in the XY plane, and is not
%orthogonal to $\hat{p}$ (unless
%$\hat{p}$ happens to lie exactly along $\hat{n}$).

 %   However, the rotation around axis $\hat{n}$ carries
%$\hat{Z}$ into $\hat{D}$, and likewise carries $\hat{p}$
%into a rotated $\hat{p}$.  This rotated $\hat{p}$
%remains perpendicular to rotated $\hat{Z}$ = $\hat{D}$.

    Since \E{x}{A} brings down a factor of spin (\eq{Ytransf}),
$<E^x_A>$ is essentially the peak value of angular momentum.
The following quantum mechanical analogy helps in
understanding the exact connection between $<\E{x}{A}>$
and $\hat{D}$.  Consider an
electron moving in a central potential and described by polar and azimuthal
coordinates $(\theta/2, \phi - \pi /2)$. (The definitions of
polar and azimuthal angles
are unorthodox but acceptable.) The angular wavefunction for the
electron, as well as the coherent state \eq{defcoh}, are both
superpositions of spherical functions $Y_{LM}$, or equivalently
rotation matrices $D(h)_{0M}$. Therefore \eq{defcoh} can serve as
a coherent state describing the angular motion of an electron,
when the position of the electron is peaked at $\hat{D}$.

    Since the angular momentum vector of the electron is
perpendicular to the orbit, we have
\begin{equation}
    0 = \sum_M \hat{D}_M (u_x)<\E{x}{M}>.
            \label{perpconstraint}
\end{equation}

    The  constraint \eq{perpconstraint} follows from the symmetry, not
the dynamics; therefore
the spin network coherent state must obey the same constraint.
The expectation value $<\E{x}{A}>$ must be a linear combination of $\hat{n}$
and $\hat{n} \times \hat{D}$.

    This section has sketched the general strategy.  The next two
sections apply it to specific cases.

\section{Action of the {\boldmath $\tilde{E}$}} \label{SecEtld}

    This section computes the action of an \Etld operator
on the coherent state.  The main results of
this section are the following: the coherent
states are approximate eigenvectors of the \Etld.
For A = $\pm 1$,

\begin{align}
    (\gamma\kappa/2)^{-1}\E{x}{A} \ket{u,\vec{p}} =&
            <L +1/2> (\hat{n}_A \cos \mu -(\hat{n}\times \hat{D})_A \sin \mu)\ket{u,\vec{p}} \nn
                &+ SC \nn
                =& <L +1/2> \hat{p}_B  \,D^{(1)}(u)_{BA} \ket{u,\vec{p}}+ SC.
\label{Eeigenval}
\end{align}
L + 1/2 is an approximation and abbreviation for $\sqrt{L(L+1)}$.
The peak value of the angular momentum is
\[
    <L +1/2> = p/t.
\]
From the last line of \eq{Eeigenval}, the direction of L is
given by the \emph{rotated} value of $\hat{p}$.  Consistent with the
electron analogy given in the last section,  rotated $\hat{p}$
is a linear combination of the two basis vectors perpendicular to $\hat{D}$.
The small correction terms SC are investigated in appendix
\ref{AppSC} and shown to be down by factors of order $1/\sqrt{L}$.

    $\hat{n}$ and $\hat{p}$ are the axes of rotation for u and H,
defined at \eqs{defu}{geqHu}.  $\mu$ is the angle
between $\hat{n}$ and $\hat{p}$, \eq{defmu}.  $\hat{D}$ is an abbreviation
for the unit vector $D^{(1)}(u)_{0A}$.

    The A = 0 component of \eq{Eeigenval} is the
\E{x}{0} introduced at \eq{defEsubZero}, where it was defined
by its action on the basic holonomy $h^{(1/2)}$.  In the present
context, if \E{x}{\pm} yields a spin matrix factor $S_{\pm}$ multiplying
each D(h) the coherent state, then \E{x}{0} may be defined as the
operator which produces a factor of $S_0$, i.e. it  multiplies
$D(h)_{0M}$ by M.
\[
    <(\gamma\kappa/2)^{-1}\E{x}{0}> \,=  \, <M_x>.
\]
$M_x$ is the U(1) quantum number needed to check U(1) invariance.  From
\eq{Eeigenval} with A = 0,
\begin{equation}
    <M_x> \,= \,<L_x> \,\hat{p}_B \,D^{(1)}(u)_{B0} .
\label{MeqHatp}
\end{equation}

    I now derive \eq{Eeigenval}.   By construction, the
D(h) matrices in the coherent state
transform simply under the action of an \Etld:

\begin{gather}
    (\gamma\kappa/2)^{-1}\E{x}{A}\ket{u,\vec{p}} =
        N\sum_{L,M}((2L+1)/4\pi) \exp[-t L(L+1)/2 \,] \nn
            \cdot D^{(L)}(h)_{0N} \,\bra{L,N} S_A\ket{L,M} \,D^{(L)}(g\dag)_{M0} \nn
            = N\sum_{L,M,R}((2 \,L+1)/4\pi) \,\exp[-t L(L+1)/2 \,] \nn
             \cdot D^{(L)}(h)_{0N} \,\bra{L,N} S_A\ket{ L,M} \,
            D^{(L)}(u\dagger)_{MR} \,D^{(L)}(H)_{R0}\nn
            = N\sum_{L}((2 \,L+1)/4\pi) \,\exp[-t L(L+1)/2 \,]\nn
             \cdot [ \,\textbf{D}^{(L)}(h) \,
             \textbf{S}_A \,
            \textbf{D}^{(L)}(u\dagger) \,\textbf{D}^{(L)}(H) \,]_{00}
                        \label{graspcoh1}
\end{gather}
On the second line I use the g = Hu decomposition,
\eq{geqHu}, plus $D(g\dag) = D(u\dag) \,D(H)$. On the last
line I have used a matrix notation to hide some indices.

    I now implement the general procedure outlined in the
previous section.   First I must
produce a factor of $D^{(1)}(u)$.  The spin
generator S in \eq{graspcoh1} is essentially a Clebsch-Gordan
coefficient, and from the rotation properties of these
coefficients, appendix \ref{Appsphcomp}, \eq{Srotn},

\[
 \mathbf{S}_A \,\mathbf{D}^{(L)}(u\dagger)=
\mathbf{D}^{(L)}(u\dagger) \,\mathbf{S}_B  \,D^{(1)}(u)_{B A}.
\]
\Eq{graspcoh1} becomes

\begin{gather}
    (\gamma\kappa/2)^{-1}\E{x}{A}\ket{u,\vec{p}} =
        N\sum_{L,M}((2 \,L+1)/4\pi) \,\exp[-t L(L+1)/2 \,]\nn
            \cdot [\mathbf{D}^{(L)}(h u\dagger) \,
            \mathbf{S}_B \,
            \mathbf{D}^{(L)}(H) \,]_{00} \,D^{(1)}(u)_{BA}\nn
            = N\sum_{L,M} ((2 \,L+1)/4\pi) \exp[-t L(L+1)/2] \,D^{(L)}(hu\dag)_{0M} \,D^{(L)}(H)_{M0}\nn
\cdot  [ \,\bra{L,M}S_B \ket{L,M-B} \,D^{(L)}(H)_{M-B,0}/D^{(L)}(H)_{M0}]\nn
           \cdot D^{(1)}(u)_{BA}.
                        \label{graspcoh2}
\end{gather}
The final three lines exhibit the desired factor of $D^{(1)}(u)$,
as well as factors of $D^{(L)}(hu\dag)_{0M} \,D^{(L)}(H)_{M0}$
which make the sum in \eq{graspcoh2} look  as much as
possible like the original coherent state.

    The sum is not exactly the original state because of
the square bracket, second
line from the end of \eq{graspcoh2}.
The Gaussian behavior of the
D(H) matrices could be used to simplify this bracket; but
is slightly more accurate to use a recurrence relation which
follows from \eq{approxCsh} of appendix
\ref{AppDH} in the $e^p \gg 1$ limit.

\begin{align}
    D^{(L)}(H)_{M\mp1,0} \cong &
    \,D^{(L)}(H)_{M,0}\,(\hat{p}_1-i\hat{p}_2)]^{\mp 1} \nn
                       &\cdot 1/\sqrt{(L\pm M)/(L \mp M + 1)}.
                        \label{DHpm1}
\end{align}
I have specialized to $p_3 = 0$.   For $\vec{p} \,$
parameterized as at \eq{defmu},
$\hat{p}_1-i\hat{p}_2 = \,\exp[-i (\beta + \mu) \,]$.  Also, the
matrix element of $S_0$ is just M, while $S_B$ for B
= $\pm 1$ is given by

 \be
    \bra{L, M} S_{\pm 1} \ket{L,M \mp 1} =\mp
             \sqrt{(L \mp M + 1)(L \pm M)/2} \label{S}
\ee
\Eq{graspcoh2} becomes

\begin{gather*}
    (\gamma\kappa/2)^{-1}\E{x}{A}\ket{u,\vec{p}} =
        N\sum_{L,M,B}((2L+1)/4\pi) \,\exp[-t L(L+1)/2 \,] \\
           \cdot  D^{(L)}(h u\dagger)_{0M} \,C[M,B] \,D^{(L)}(H)_{M 0} \,
                \exp[ \,i B(\beta + \mu) \,] \,D^{(1)}(u)_{B A};
\end{gather*}
\bea
                 C[M,B=0] &=& M;\nn
                 C[M,B =+1] &=& -(L+M)/\sqrt{2}; \nn
                 C[M,B = -1]&=& (L-M)/\sqrt{2}.
                  \label{defC}
\eea

   Using appendix \ref{AppD1u}, I replace the $D^{(1)}(u)_{B A}$
by the geometrically more
transparent quantities $\hat{n}$,
$\hat{D}$, and $\hat{n}\times \hat{D}$.   The row $D^{(1)}(u)_{0
A}$ is just the vector $\hat{D}$ introduced at \eq{defDhat}.
The rows $D^{(1)}(u)_{\pm 1, A}$ may be replaced by  linear
combinations of the unit vectors $\hat{n}$ and
$\hat{n}\times \hat{D}$, using \eqs{ncomp}{nxVcomp} in
appendix \ref{AppD1u}.  \Eq{defC} becomes

\begin{multline}
    (\gamma\kappa/2)^{-1}\E{x}{A}\ket{u,\vec{p}} =
         N\sum_{L,M}((2 \,L+1)/4\pi) \,\exp[-t L(L+1)/2 \,]\\
           \cdot D^{(L)}(h u\dagger)_{0M} \,D^{(L)}(H)_{M0} \\
          \cdot  \{(L \cos \mu +i M \sin \mu)\hat{n}_A
            +(- L \sin \mu +i M \cos \mu) (\hat{n}\times \hat{D})_A \\
            +M \,\hat{D}_A \}.
                \label{graspcoh3a}
\end{multline}

    The third and final step of the general procedure
invokes the Gaussian nature of the factor D(H).
From appendix \ref{AppDH}
\begin{multline*}
    D^{(L)}(H)_{M 0}\exp[-t L(L+1)/2] \cong \\
                \sqrt{t/\pi}\exp[-t(L+1/2-p/t)^2/2 \,] \\
             \cdot   \sqrt{1/[(L+1/2)\pi]} \,\exp\{-M^2/[2(L+1/2) \,]\} \,f(p,t).
\end{multline*}
This suggests expanding the
factors of L +1/2 and M in the curly bracket around
their means, $<L +1/2> $= p/t  and $<M>$ = 0.  Equivalently,
in \eq{graspcoh3a} set
\begin{align}
    L +1/2 \,& = p/t + (L + 1/2 - p/t) \nn
            & :=  \: <L +1/2> + \,\Delta L;\nn
            M & = \Delta M,
\end{align}
and keep out to first order in $\Delta L$ and $\Delta M$.
The leading
terms in the expansion come from the terms proportional to L + 1/2
in the curly bracket, equation \eq{graspcoh3a}.
When L is expanded as L + 1/2 = $<\mathrm{L + 1/2}> + \Delta \mathrm{L}$  one finds a leading
term proportional to
\[
        <L + 1/2> (\hat{n} \cos \mu -\hat{n}\times \hat{D})_A \sin \mu).
\]
This expression is a constant and may be taken out of the sum,
leaving behind just the
original coherent state.  We get the desired result \eq{Eeigenval}
for the leading term.

    The remaining terms in the expansion around peak M and L
all involve $\Delta X$, X = L or M.  These small correction (SC)
terms are studied in appendix
\ref{AppSC}.  The sums over L and M are approximated by integrals,
and the SC terms are then found to be down by factors of
order $\sigma_X/<L>$, where $\sigma_X$ is the standard deviation for the
Gaussian distribution of X.

    Since $\sigma_L = 1/\sqrt{t}$ depends on t, from
\eq{DHforp3eq0}, evidently the "arbitrary" width
t is not so arbitrary after all.  I show in the appendix that
the most reasonable value, the one which minimizes all SC terms
to the same extent, is t = order $1/<\mathrm{L}>$.

\section{Matrix elements of the holonomy} \label{SecHolonomy}

    This section computes the action of the holonomy $h^{(1/2)}$
on the coherent state.  Presumably the
Hamiltonian will be a function of this matrix; but for
calculations it is more convenient to use the linear
combinations $D^{(1)}(h)$, \eqs{heqY}{YeqD}, rather than the
$h^{(1/2)}$.  The coherent
state is an expansion in spherical harmonics $D^{(L)}(h)$, and using
the matrix $D^{(1)}(h)$ allows one to
invoke identities for simplifying the product of two D's.  Of
course once it is shown that the peak value of $D^{(1)}(h)$
is $D^{(1)}(u)$, it follows that the peak value of $h^{(1/2)}$
is just $u^{(1/2)}$.

    Some preliminary qualitative remarks may help the reader
once again avoid burial under detail.  $D^{(1)}(h)$ multiplies the
$D^{(L)}(h)$ contained in the coherent state; this quadratic
combination of D's may be rewritten as a linear combination
of $D^{(L\pm 1)}$ and $D^{(L)}$, using Clebsch-Gordan coefficients
and the angular momentum addition rules.  (Actually, the CG
coefficients forbid any $D^{(L)}$.  Absence of $D^{(L)}$ also
follows from parity conservation.)  The coherent state is clearly
not an exact eigenfunction of $D^{(1)}(h)$, but the
Gaussian nature of D(H) comes to the rescue.  The states
involving $D^{(L\pm 1)}$ are shown to be coherent states peaked at L$ \pm$ 1,
with sufficiently large standard deviation that these coherent
states are indistinguishable from the original coherent state
with peak at L.

    After these qualitative remarks,
we are ready for the calculation.

\begin{multline}
    D^{(1)}(h)_{0A}\ket{u,\vec{p}} = N\sum_{L,M}((2 \,L+1)/4\pi) \,\exp[-t L(L+1)/2 \,]\\
         \cdot \sum_{L\pm 1} D^{(L\pm 1)}(h)_{0M} \,\braket{L \pm1 ,0}{L,0;1,0} \\
          \cdot  \braket{L \pm 1,M}{ L,M-A;1,A} \\
          \cdot D^{(L)}(u\dagger)_{M-A,N} \,D^{(L)}(H)_{N0}
                \label{hcoh1}
\end{multline}
On the right I have rewritten $D(g\dag) = D(u\dagger) \,D(H)$, from
\eq{geqHu}; and I have used the formula for combining two D(h) matrices:

\begin{multline*}
    D^{(L)}(h)_{0,M-A} \,D^{(1)}(h)_{0A} = \sum_{L\pm 1} D^{(L\pm 1)}(h)_{0M} \\
             \cdot \braket{L \pm1 ,0}{L,0;1,0}
            \braket{L \pm 1,M}{ L,M-A;1,A}.
\end{multline*}
This formula is a special case of the general relation for
coupling two D's, \eq{DDiden}, appendix \ref{Appsphcomp}.  As mentioned earlier, the value
L is excluded from
the sum on the right; and  the sum over L
involves D's with $L \,\rta \, L \pm 1$.  For the moment,
I postpone an investigation
of the $ D^{(L\pm 1)}$  and continue with the general
procedure outlined in section \ref{SecGenForm}.

    I introduce  the basis vectors $D^{(1)}(u)$ by again invoking
\eq{DDiden}.

\begin{multline*}
    \braket{L \pm 1,M}{ L,M-A;1,A} \,D^{(L)}(u\dagger)_{M-A,N} \\
    = D^{(L \pm 1)}(u\dagger)_{M,N+B} \,\braket{L \pm 1,N+B}{ L,N;1,B} \,
            D^{(1)}(u)_{BA}.
\end{multline*}
I insert this into \eq{hcoh1}, and relabel $N \rightarrow
N-B$.

\begin{multline}
     D^{(1)}(h)_{0A}\ket{u,\vec{p}} \\
        = N\sum_{L,N}((2L+1)/4\pi)\exp[-t L(L+1)/2 \,]\\
         \cdot \sum_{L\pm 1} D^{(L\pm 1)}(hu\dagger)_{0N} \,D(H)_{N0}\\
       \cdot  [ \,\braket{L \pm 1,0}{L,0;1,0} \,\braket{L \pm 1,N}{ L,N-B;1,B} \\
        \cdot  D^{(L)}(H)_{N-B,0} \,/D(H)_{N,0} \,]  \,D^{(1)}(u)_{BA} .
            \label{hcoh2a}
\end{multline}
I have multiplied and divided by $D(H)_{N0}$, so that
the second and third lines resemble the original coherent state,
except for the change L $\rta$ L $\pm$ 1.

    I must now evaluate the square bracket on the last two lines.  As for
\E{x}{A}, I use a recurrence relation which follows from \eq{approxCsh},
appendix \ref{AppDH}, to evaluate the ratio
\[
    D(H)_{N-B,0}/D(H)_{N0}.
\]
I relabel N $\rta$ M, and use the results of appendix \ref{AppD1u}
to replace $D^{(1)}(u)$ by $\hat{n}$, $\hat{D}$, and
$\hat{n}\times \hat{D}$.

\begin{multline}
    D^{(1)}(h)_{0,A}\ket{u,\vec{p}} \\
        = N\sum_{L,M}(2L+1)/(4\pi)\exp[-t L(L+1)/2 \,] \\
         \cdot \sum_{L\pm 1} D^{(L\pm
            1)}(hu\dagger)_{0M} \,D^{(L)}(H)_{M0}\\
         \cdot (c(L \pm 1,D)\hat{D}_A + c(L \pm 1,n)\hat{n}_A \\
         + c(L \pm 1, \hat{n}\times \hat{D})(\hat{n}\times \hat{D})_A).
                \label{hcoh2b}
\end{multline}
The coefficients c are

\begin{align}
    c(L+1,\hat{D}) &= \braket{L + 1,0}{L,0;1,0}\braket{L+1,M}{ L,M;1,0};\nn
    c(L + 1,\hat{n}) &= \braket{L + 1,0}{L,0;1,0} \nn
        &\cdot \sum_{\pm} ((\mp 1/\sqrt{2})\braket{L + 1,M}{ L,M \mp 1;1,\pm 1} \nn
        &\cdot \exp(\pm i\mu) \sqrt{(L \pm M)/(L \mp M + 1)}); \nn
    c(L + 1, \hat{n}\times \hat{D}) &= i \braket{L + 1,0}{L,0;1,0} \nn
        &\cdot \sum_{\pm} ((- 1/\sqrt{2})\braket{L + 1,M}{ L,M \mp 1;1,\pm 1} \nn
        &\cdot \exp(\pm i\mu) \sqrt{(L \pm M)/(L \mp M + 1)}).
                \label{cdef1}
\end{align}
For the case L+1 $\rightarrow$ L-1, replace $\bra{L+1,M}\mbox
{and}\bra{L+1,0}$ by $\bra{L-1,M}\mbox{and}\bra{L-1,0}$.  After
some work with a table of 3J symbols, one finds

\begin{align}
    c(L+1,D) &= \sqrt{(L+1)^2 - M^2}/(2L+1);\nn
    c(L-1,D) &= \sqrt{L^2 - M^2}/(2L+1);\nn
    c(L + 1,n) &= -[ \sqrt{(L+1+M)/(L+1-M)}(M)/(2L+1) \,]\cos \mu \nn
               &    -i[\sqrt{(L+1+M)/(L+1-M)}(L)/(2L+1)\,]\sin \mu ;\nn
    c(L-1,n) &= i \,[ \,\sqrt{L^2 - M^2}/(2L+1) \,]\sin \mu;\nn
     c(L + 1, \hat{n}\times \hat{D})&=- i[\sqrt{(L+1+M)/(L+1-M)}(L)/(2L+1) \,]\cos \mu \nn
                 &    + [ \sqrt{(L+1+M)/(L+1-M)}(M)/(2L+1) \,]\sin \mu;\nn
     c(L-1,\hat{n}\times \hat{D})&= i \,[ \,\sqrt{L^2 - M^2}/(2L+1) \,] \,\cos \mu.
                    \label{cdef2}
\end{align}

    Continuing with the general procedure, I expand
the L and M dependence of the c's. I replace
\[
    L \rta <L > + \Delta L; \quad M \,\rta \,\Delta M.
\]
and keep terms in \eq{cdef2} out to linear in $\Delta X $, X = L
or M.  I have dropped the 1/2 in $<L+1/2>$ to simplify formulas.

\bea
    c(L+1,D) &\cong& <(L+1)/(2L+1)>;\nn
    c(L-1,D) &\cong& <L/(2L+1)>;\nn
    c(L + 1,n) &\cong& - (M/<L>) \cos \mu
                    -i\sin \mu <L/(2L+1)>;\nn
    c(L - 1,n) &=& i\sin \mu <L/(2L+1)>;\nn
    c(L + 1,\hat{n}\times \hat{D}) &\cong&
                    -i <L/(2L+1)>\cos \mu
                    + (M/<L>)\sin \mu ;\nn
    c(L - 1,\hat{n}\times \hat{D}) &\cong&
                    +i <L/(2L+1)>\cos \mu.
        \label{cdef3}
\eea
Factors of L inside brackets $<>$ should be interpreted as $<L>$;
terms of order $\Delta X/<L>$ have been kept, but not
terms of order $\Delta X/<L>^2$ or $(\Delta X/<L>)^2$.
After inserting this expansion into \eq{hcoh2b}, that equation
becomes

 \begin{eqnarray}
    D^{(1)}(h)_{0A}\ket{u,\vec{p}} &\cong& N\sum_{L,M}((2L+1)/4\pi) \exp[-t
                                    L(L+1)/2 \,]\nn
       & &\cdot \{(1/2) \,[D^{(L+1)}(hu\dagger)+D^{(L-1)}(hu\dagger) \,]_{0M} \,\hat{D}_A \nn
        & &    + D^{(L+1)}(hu\dagger)_{0M} \,(M/<L>) \nn
        & & \cdot [-\cos \mu \hat{n}_A +\sin \mu (\hat{n}\times \hat{D})_A] \nn
        && \quad + \,\mbox{small terms}; \nn
    \mbox{small terms \,} &= &    -i (1/2) \,[ \,D^{(L+1)}(hu\dagger)-D^{(L-1)}(hu\dagger) \,]_{0M} \nn
       & &       \cdot [ \,\sin \mu \,\hat{n}_A + \cos \mu \,(\hat{n}\times \hat{D})_A] \nn
        & & \quad +   \, \mbox{order} \,((M/L)^2,1/L)\} \,D^{(L)}(H)_{M0} .
                \label{hcoh3}
\end{eqnarray}
From \eq{DHforp3eq0}, $\sigma_M = \sqrt{L}$.  I anticipate that
\[
    M/<L> =  \mbox{order}\, \sigma_M/<L> = \mbox{order}\, 1/\sqrt{<L>}.
\]
Therefore I am keeping order
$1/\sqrt{<L>}$ but neglecting
order $(M/L)^2$ = order 1/L.  Because of this, when going from \eq{cdef3}
to \eq{hcoh3}, it is legitimate to replace $<L/(2 \,L+1)>$ by 1/2, etc..

    The sums involving $D^{(L\pm 1)}$ have not gone away.  I must now
investigate these sums.

\subsection{The kets  $\ket{\mathbf{L\pm 1}}$}

    Equation \eq{hcoh3} involves the two new kets,

\begin{align}
        \ket{L\pm 1} &= N(L\pm 1)\sum_{L,M}((2 \,L+1)/4\pi) \exp[-t L(L+1)/2 \,]\nn
              & \, \cdot  D^{(L\pm 1)} (hu\dagger)_{0M} \,D^{(L)}(H)_{M0}.
                    \label{defpmdeltL}
\end{align}
This subsection shows that these kets are Gaussian
distributed in L, but with peak values of L shifted by one unit,
from $<L +1/2>$ to $<L +1/2>\pm 1$.   This result has a corollary.  In
\eq{hcoh3} the
linear combination with the upper sign, $D^{(L+1)}(hu\dagger)+D^{(L-1)}(hu\dagger)$,
is a sum of two Gaussians in L with slightly different means, but large standard
deviations $\sigma_L = \sqrt{1/t}$, 1/t = \mbox{order $<L>$}.  Therefore the two Gaussians
strongly overlap with each other, and the sum strongly resembles
the Gaussian of the original coherent state $\ket{u,\vec{p}}$.
The remaining combination $D^{(L+1)}(hu\dagger)-D^{(L-1)}(hu\dagger)$ is the difference
between two closely similar Gaussians, therefore is very small.  Proof that
the difference is small requires a detailed calculation of its normalization
constant, and I do this calculation in appendix \ref{AppSC}.

   The states defined at \eq{defpmdeltL} have peak angular
momentum shifted by one unit. Proof: in
\eq{defpmdeltL}, the D(H) factor is easier to approximate, while
$D(hu\dagger)$ is harder.  Therefore relabel $L \pm 1 =
\tilde{L}$, then drop tildes, in order to make the  $D(hu\dagger)$
factor equal to the corresponding factor in the original coherent
state $\ket{u,\vec{p}}$: under the relabeling,
$D^{(L\pm1)}(hu\dagger) \,\rta \,D^{(\tilde{L})}(hu\dagger) \,\rta \,
D^{(L)}(hu\dagger)$.  The
relabeling changes the exponential $\exp[-t L(L+1)/2]$
significantly; and it produces negligible changes in
$(2L+1)/4\pi\cong (2\bar{L}+1)/4\pi + \mbox{order}(1/L)$

    The new
D(H) factor is $D^{(L)}(H) \,\rta \,D^{(\tilde{L}\mp1)}(H) \,\rta \,
D^{(L\mp1)}(H)$. This matrix differs from the corresponding matrix in
$\ket{u,\vec{p}}$, but is easy to relate to that matrix.  I group
together the two factors in \eq{defpmdeltL} that change
significantly, and use \eq{DHforp3eq0}.

\begin{multline}
    \exp[-t (L\mp 1)(L \mp 1 + 1)/2] \,D^{(L\mp1)}(H)_{M0} \\
        \cong \exp\{-t [L\mp1+1/2 \,]^2/2+t/8 \, \}
                \exp[-i M \beta \,] (\exp(p/2)/2)^{2L \mp 2} \\
          \cdot  \frac {(2L \mp 2)!}
            {\sqrt{(L \mp1)!^2(L \mp1+M)!(L\mp1-M)!}}\\
        \cong \exp\{-t[L\mp 1+1/2 -p/t \,]^2/2 +p^2/2t-p/2 \,\}\\
    \cdot    \exp\{-M^2/[2(L \mp 1+1/2) \,] \,\}/\sqrt{\pi(L \mp 1 +1/2)}\\
        \cong \exp\{-t[L \mp1+1/2 -p/t \,]^2/2 +p^2/2t-p/2 \,\} \\
        \exp\{-M^2/[2(L +1/2) \,]\}/\sqrt{\pi(L+1/2)}.
        \label{nearDH}
\end{multline}
The factorials were replaced by a Gaussian using \eq{GaussianM}.
Now compare the Gaussian in \eq{nearDH} to the Gaussian in the original
coherent state, \eq{DHforp3eq0}.  The two
Gaussians in L are the same except for a shift in peak value.

\be
    \mbox{new}<L+1/2> = p/t \pm 1
                    = \mbox{original} \,<L+1/2> \pm 1.
            \label{newL}
\ee
$\Box$

    Now separate \eq{hcoh3} into a leading  term plus small corrections.
\Eq{hcoh3} involves the linear combinations
\[
    (1/2) \,[ \,D^{(L+1)}(hu\dagger)\pm D^{(L-1)}(hu\dagger) \,],
\]
therefore linear combinations
\[
    (1/2) \,(\ket{L+1}\pm \ket{L- 1})
\]
of the kets just investigated.
Therefore, inside a sum over L,

\begin{gather}
    (1/2)[ \,D^{(L+1)}(hu\dagger)
        +D^{(L-1)}(hu\dagger) \,] \,D^{(L)}(H) \nn
        \cong D^{(L)}(hu\dagger) \,D^{(L)}(H);\nn
    (1/2)( \,\ket{L+1}+ \ket{L- 1} \,) \,\cong \, \ket{u,\vec{p}}.
\label{sumeqcoh}
\end{gather}
The linear combination with the upper sign will yield a dominant term.

    The $\hat{n}$ terms in equation \eq{hcoh3} involve
a factor of \mbox{M/$<L>$}, and the discussion of appendix \ref{AppSC} shows these
terms are down by $1/\sqrt{L}$, as expected
because  $D^{(L)}(H)$ is Gaussian in M.  Therefore only one term is dominant,
the one proportional to $\hat{D}_A$.
From \eq{defDhat} $\hat{D}_A$ is also $D^{(1)}(u)_{0A}$;
and we get

\be
    D^{(1)}(h)_{0A}\ket{u,\vec{p}} = D^{(1)}(u)_{0A} \,
            ( \,\ket{L+1} + \ket{L-1} \,)/2 + SC.
            \label{hEigenval}
\ee

    I have not
replaced the linear combination
\[
    (1/2) \,( \,\ket{L+1}+ \ket{L- 1} \,)
\]
by $\ket{u,\vec{p}}$, even though the three states
overlap strongly.
The linear combination   $(1/2) \,( \,\ket{L+1}+ \ket{L- 1} \,)$
has the same \Etld eigenvalues as  $\ket{u,\vec{p}}$.  However,
$(1/2) \,( \,\ket{L + 1} + \ket{L - 1} \,)$ does not behave
like $\ket{u,\vec{p}}$ under the action of a holonomy:

\begin{multline*}
    D^{(1)}(h)_{0,A} \,(1/2) \,( \,\ket{L + 1} + \ket{L - 1} \,)\\
        \cong (1/4) \,( \,\ket{L + 2} + 2\ket{L}+ \ket{L - 2} \,),
\end{multline*}
where $\ket{L+n}$ has peak L shifted by n.
I am not yet sure whether these changes in peak L have any
significance, and for now, I prefer to keep the shifts in L explicit. Note
if the Hamiltonian contains a product of n holonomies, then
the original coherent state will be changed into a linear combination
of coherent states with peak L shifted by up to $\pm n$ units.

 \section{The  commutator [h,V]} \label{SecVhcomm}

    At this point it might seem we are finished. The basic
building blocks for the Hamiltonian are the holonomy and \Etld;
formulas of the previous sections should be adequate to evaluate most
matrix elements.

    There is one exception, however.  The formalism makes extensive
use of the volume operator \Vthree.  For the planar case this
operator factors:

\bea
        (\Vthree)^2 &=& sgn \,\E{z}{Z} \, \epsilon_{ZAB} \, \E{x}{A} \, \E{y}{B} \nonumber \\
                &:=& sgn \,\E{z}{Z} \, \Etwo.
    \label{V3}
\eea
sgn is the sign of the determinant of the \Etld.
One constructs extrinsic curvatures by commuting this operator with the Hamiltonian.
Therefore the formalism contains matrix elements such as
\begin{eqnarray*}
    \bra{u,\vec{p}}[h^{(1/2)},\Vthree]\ket{u,\vec{p}} &=&
        \bra{u,\vec{p}}h^{(1/2)}\ket{n}\bra{n}\Vthree\ket{u,\vec{p}}\\
        && - \bra{u,\vec{p}}\Vthree\ket{n}\bra{n}h^{(1/2)}\ket{u,\vec{p}},
\end{eqnarray*}
where $h^{(1/2)}$ is an SU(2) holonomy coming from the Hamiltoninan.  This
commutator is a small difference between two large terms.  The contribution
from the leading term, $\ket{n} = \ket{u,\vec{p}}$, cancels
out of the difference, and it is very hard to evaluate this
commutator using only the results of previous sections.

    Thiemann and Winkler have shown that, in the classical limit,
the quantum mechanical commutator equals the classical Poisson
bracket \cite{GCSIII}.
\be
    \bra{u,\vec{p}}[ \,O_1,O_2 \,]\ket{u,\vec{p}}/i\hbar
            =  \{O_1,O_2\}(u,\vec{p})
            \label{TWtheorem}
\ee
On the right, one computes the Poisson bracket, treating the
operators as classical fields, then evaluates the result at
the peak values for the coherent state.  Since the Poisson
bracket is no longer a small difference between two large
terms in general, it can be evaluated
using the formulas of previous sections.

    While  this is not the place to construct a full
Hamiltonian, some discussion of its likely,properties
is appropriate, in order to decide which [V,h] commutators are
relevant.  \Vthree must be commuted with the Euclidean
Hamiltonian, which contains the following terms (in a
field theoretic formulation, before discretization on a spin
network)

\bea
   && F^Z_{xy} \,\Etwo \,/ \,\Vthree;\nn
    && F^A_{zx} \,\E{z}{Z} \,\E{x}{B} \,\epsilon_{ZBA}/ \,\Vthree;\nn
   &&  F^A_{zy} \,\E{z}{Z} \,\E{y}{B} \,\epsilon_{ZBA}/ \,\Vthree.
    \label{F}
\eea
A,B = X,Y only.  In spin network theory for
the planar case, these terms become commutators with spin
1/2 holonomies along the x, y, and z edges.
\bea
   &&2 \,F^Z_{xy} \,Tr\{\sigma_Z \,h_z [ \,h_z^{-1},\Vthree \,]\};\nn
    &&2 \,F^A_{zx} \,Tr\{\sigma_A \,h_y [ \,h_y^{-1},\Vthree \,]\};\nn
    &&2 \,F^A_{zy} \,Tr\{\sigma_A \,h_x [ \,h_x^{-1},\Vthree \,]\}.
    \label{HeqVhcomm}
\eea

    \Eq{HeqVhcomm} is not yet in its final spin network
form.  Each $F_{ij}$ must be replaced by a product
of four holonomies encircling the four sides of the area ij.
In general a product of four spin 1/2 holonomies will contain all the Pauli
matrices.  For instance, the holonomies replacing $F^Z_{xy} \,\sigma_Z$
could contain $\sigma_A$ matrices, A $\neq$ Z, or even the 2x2
unit matrix.   Fortunately, there exist several possible holonomy
products which reduce to the same $F_{ij}$ in a field theory
limit.  (For example,
proceed counterclockwise \textit{vs.} clockwise around the area.)
One can choose a linear combination of the possibilities which is
pure $\sigma_Z$ or pure $\sigma_A$, as needed to reproduce
the classical results.  In what follows I will assume this has been
done, so that \emph{the only traces required are those shown in \eq{HeqVhcomm}},
the $\sigma_Z$ trace for $h_z$ and the $\sigma_A$ traces, A
transverse, for $h_x$ and $h_y$.

\subsection{Holonomies along z}

    In the planar case it is possible to evaluate the
$[ \,h_z^{-1},\Vthree \,]$ commutator exactly, if the z dependence
of the coherent state is chosen to be an eigenfunction of \E{z}{Z}.
It is helpful to derive this
exact result, in order to clarify some issues regarding \eq{TWtheorem}.

    Consider the $[ \,h_z,\Vthree \,]$ commutator, with
the z dependence at the nth vertex chosen to be eigenfunctions of
\E{z}{Z}.  This
vertex is the terminus for two holonomies directed along
z, one incoming and one outgoing.  Therefore the wavefunction at
vertex n contains the product
\bea
    h(z,n) &:=& \exp[\,i \int_{n-1}^n \A{Z}{z} \,S_Z \,dz \,] \,
                \exp[\,i \int_n^{n+1} \A{Z}{z} \,S_Z\, dz \,] \nn
           &=& \exp[\,i \int_{n-1}^n \A{Z}{z} \,M_i \,dz \,] \,
            \exp[\,i \int_n^{n+1} \A{Z}{z} \,M_f \,dz \,].
            \label{defhzn}
\eea
The total wavefunction is h(z,n) $\ket{u_x,\vec{p}_x} \,\ket{u_y,\vec{p}_y}$.
When  $h_z \,[h_z^{-1},\Vthree]$ acts on this wavefunction, the
$\sqrt{\mid \Etwo \mid}$ factor in $\Vthree = \sqrt {\mid \E{z}{Z} \,\Etwo \mid}$
commutes through to the $\ket{u_x,\vec{p}_x}\,\ket{u_y,\vec{p}_y}$ factor, while the
$\sqrt{\mid \E{z}{Z} \mid}$ factor acts on
the product of the $h_z^{-1}$ factor in the commutator and the h(z,n)
factor in the state.

    To compute the action of $\sqrt{\mid \E{z}{Z} \mid}$, we must first compute
the action of \E{z}{Z} (without the square root) then take the square root of
this operator (which is easy, since the operator is diagonal).  The action
of \E{z}{Z} will be a sum of two amplitudes, since h(z,n) contains two factors,
one ingoing and one outgoing.

\begin{multline}
       [ \,h_z^{-1}, \E{z}{Z} \,] \,h(z,n)\\
        = [\exp\{-i \int_n^{n+1}\A{Z}{z}(\pm 1/2) \,dz\},\E{z}{Z}]\, h(z,n)\\
        = (\gamma\kappa/2) h_z^{-1} h(z,n)[(M_f+M_i) -(M_f +M_i \mp 1/2)].
        \label{hEzcomm}
\end{multline}
The amplitude for the outgoing (f for "final") factor in h(z,n) gives rise to the
$M_f$ terms in the square bracket, while the $M_i$ terms come from the
incoming (i for "initial") factor.  The $\pm$ refers to the $[\pm,\pm]$ diagonal
element of the matrix $h_z^{-1}$.

    Earlier in this section I stated that it was easy to
take the square root of the
\E{z}{Z} operator, because this operator is diagonal.  On the
contrary: the literature contains at least two recipes
for extracting the square root of this operator.  At present
it does not seem
possible to distinguish between the two approaches, using general
principles; and I need to state which approach I am using here.

    When more than three edges terminate at a given vertex, the
$(\mbox{volume})^2$
operator grasps each triplet of edges in turn, generating a series
of amplitudes, one for each triplet.  Similarly, when two z holonomies
terminate at a vertex, the \E{z}{Z}  operator generates two amplitudes.
To parallel a  terminology from classical optics: should one add
first, then take the square root?  Or take the square root first,
then add?  I.e. should one
add the amplitudes from each triplet (or z holonomy), then take
the square root of
the magnitude of the result; or should one take the square
root of the magnitude of each amplitude first, then add?
The literature contains advocates for both
"add first" \cite{ALvolumeI,ALvolumeII} and "take the square root first"
\cite{RSvolume,DePRvolume} choices.  For a
discussion of the distinct regularization schemes leading to each
choice see reference \cite{ALvolumeII}.  Presumably no final choice between
the schemes can be made until the two choices have been checked
in applications.

    In this paper I adopt the "add first" choice, for the following
(non-rigorous) reason.  Consider a four-valent vertex such that
two of the edges meeting at the vertex are tangent, one ingoing and
one outgoing.  Rotate the gauge at the vertex so that the holonomies
along these edges are pure $S_Z$, with $S_Z$ eigenvalues $M_f$ and
$M_i$ for outgoing and ingoing vertices, respectively.  (This is
of course exactly what we have already, in the planar case, due to
the gauge-fixing.)  The
vertex has two holonomies,
\[
    \exp[ \,i\int_{n}^{n+1} M_f \,\A{Z}{i}dx^i \,]; \quad
    \exp[ \,i\int_{n-1}^{n} M_i \,\A{Z}{i}dx^i \,].
\]
In the "add first" prescription, the contributions to
$(\mbox{volume})^2$ from these two holonomies are added,
giving a factor of $(M_f + M_i)/2$.  The factor
of 1/2 comes from the integrations over half a delta function.
The volume is proportional to the square root
\[
    \sqrt{\mid M_f + M_i \mid/2} .
\]
In the "square root first" prescription, the corresponding
factor would be
\[
    \sqrt{\mid M_i \mid /2} + \sqrt{\mid M_f \mid /2}.
\]
In the special case $M_f = M_i$, it is possible to view these
two holonomies as a single
holonomy passing through the vertex.  When this is grasped by
the  $(\mbox{volume})^2$ the contribution is proportional
to $M_i$ with no 1/2, and the volume is proportional to
$\sqrt{\mid M_i \mid}$.   This result for one holonomy equals
the $M_f = M_i$ limit of the result for two different holonomies,
only if the "add first" prescription is used.

    That prescription gives the following result for the commutator.
\bea
    [ \,h_z^{-1}, \sqrt{\eta \,\E{z}{Z}} \,] \,h(z,n)
        &=& \sqrt{(\gamma\kappa/2)} \, h_z^{-1} \,h(z,n)[\sqrt{ \,\mid M_f+M_i\mid \,} \nn
        & &   -\sqrt{ \,\mid M_f +M_i \mid \mp \,\eta/2 \,}],
            \label{hrootEzcomm}
\eea
where $\mid \E{z}{Z}\mid \,= \eta \,\E{z}{Z}$ and $ \eta = \pm 1$ is the phase
of the eigenvalue $ M_f + M_i $.  That is, I "add first"  the $ M_f$ and
$M_i $ contributions, then take the square root.  Note the
$ M_f+M_i$ and $M_f +M_i \mid \mp \,1/2$ parentheses in \eq{hEzcomm}
have not been combined or averaged; they come from two
separate applications of the volume operator.

    From \eq{HeqVhcomm} the result \eq{hrootEzcomm} should be inserted
into the trace
\bea
    T(Z,z) \,h(z,n) &:=& Tr\{ \,\sigma_Z \,h_z \,[ \,h_z^{-1},\sqrt{\eta \,\E{z}{Z}}] \,\} \,h(z,n) \nn
           &=& \Sigma_{\pm}\{(\pm 1)\sqrt{(\gamma\kappa/2)} \,[\sqrt{ \,\mid M_f+ M_i\mid \,} \nn
            & &    -\sqrt{ \,\mid M_f +M_i \mid \mp \,\eta/2} \,] \,\} \,h(z,n) \nn
           &=& \sqrt{(\gamma\kappa/2)}[ -\sqrt{\mid M_f +M_i \mid - \,\eta/2} \nn
            & &         +\sqrt{\mid M_f +M_i \mid + \,\eta/2}] \,h(z,n).
                     \label{defTZz}
\eea

    \Eq{defTZz} is an exact result.  In the limit of large quantum numbers
$\mid M_f+M_i\mid \gg 1/2$, the radicals can be expanded, yielding
\be
    T(Z,z) \,\rta \,\sqrt{(\gamma\kappa/2)}[ \,\eta/(2\sqrt{ \,\mid M_f+M_i\mid \,}) \,].
        \label{limTZz}
\ee
This approximate result should be compared to the result from classical field
theory (or from the theorem, \eq{TWtheorem})
\bea
    T(Z,z) \,h(z,n) &\rta& Tr\{\sigma_Z (1) \,[ \,1-i \A{Z}{z}S_Z dz,\sqrt{\eta \,\E{z}{Z}}] \,\}h(z,n)\nn
            &=& Tr\{\sigma_Z (1) \,[ \,S_Z \,dz (\delta/\delta \E{z}{Z})\sqrt{\eta \,\E{z}{Z}}] \,\}h(z,n)\nn
            &=& (\eta/2)(\gamma\kappa/2)[1/\sqrt{\eta \,\E{z}{Z}} \,] \,h(z,n) \nn
            &=& (\eta/2) \sqrt{(\gamma\kappa/(2\mid M_f+M_i\mid)} \,h(z,n).
            \label{fieldthTZz}
\eea
This result agrees with the large quantum number limit, \eq{limTZz}.

    This agreement helps in understanding a rather puzzling feature of
the theorem, \eq{TWtheorem}.  The $\sqrt{\eta \,\E{z}{Z}}$
operator is not distributive.  When acting on a product of two functions
of holonomies,
\[
    \sqrt{\eta \,\E{z}{Z}} \,[ \,f_1(h) \,f_2(h) \,] \neq f_2(h) \sqrt{\eta,\E{z}{Z}} \,[ \,f_1(h) \,]
                    + f_1(h) \sqrt{\eta \,\E{z}{Z}} \, [ \,f_2(h) \,].
\]
This can be checked readily by choosing the $f_i$ to be eigenvectors of \E{z}{Z}.
In \eq{TWtheorem} suppose one takes $O_1 = \sqrt{\eta \,\E{z}{Z}}$, $O_2 = f_1$.
The theorem seems to state that the action of $O_1$ on $f_1$ is independent of
the state on which the commutator acts (is independent of $f_2$).  One might be tempted to
question this result, given the non-distributive
character of $\sqrt{\eta \,\E{z}{Z}}$.  However, the theorem does predict
the correct classical limit, equation \eq{fieldthTZz}, in a case where we
are working from the exact quantum result.

   If the z holonomies at the vertex are coherent states, rather
than eigenfunctions of \E{z}{Z}, the formulas of this section still work,
with $ M_f$ and $M_i $ replaced by their peak values $ <M_f>$ and
$<M_i> $.

    When the eigenvalues  $\mid M_i \mid$ of $\mid \E{z}{Z} \mid$ are large,
naively one might expect
$[h_z^{-1},\sqrt{\mid \E{z}{Z}\mid} \,]$ to
be order the eigenvalue of $\sqrt{\mid \E{z}{Z}\mid}$,
$\sqrt{ \,\mid M_i + M_f \mid \,}$; but instead it is of order
$1/(2\sqrt{ \,\mid M_f+M_i\mid \,})$, which is the derivative
of $\sqrt{ \,\mid M_i + M_f \mid \,}$.  In the next subsection we will find
that this commutator = derivative structure
holds also for the commutators of transverse holonomies with the
volume.

\subsection{Holonomies along x,y}

    Now consider commutators involving $h_a, \mbox{a = x,y}$.  As before,
\Vthree factors into \Vtwo  times
$\sqrt{\mid \E{z}{Z}\mid}$; but now the
$\sqrt{\mid \E{z}{Z}\mid}$ factors out of the commutator.
We are left with the following trace, which acts on the coherent ket.

\be
    T(A,a) = Trace \,( \,\sigma_A \,h_a \,[ \,h^{-1}_a,\Vtwo \,] \,).
        \label{defT}
\ee
Since the case a = y may be obtained from results for a = x by a
relabeling, it suffices to compute this trace for a = x.

    The theorem, \eq{TWtheorem}, gives the commutator in T(A,a).
\bea
    [ \,h^{-1}_x,\Vtwo \,] &=& i [\,\delta (h^{-1}_x)/\delta \A{B}{x} \,]
           [ \,\delta (\Vtwo)/\delta \,\E{x}{B} \,](\gamma\kappa) \nn
            &=& (\gamma\kappa/2)[ \,S_B, h^{-1}_x \,]_+ [ \,\zeta \,\E{y}{C} \,\epsilon^{BC}/2\Vtwo \,],
            \label{hVPB}
\eea
where \Vtwo  is $\sqrt{\zeta \,\E{x}{B} \,\E{y}{C} \,\epsilon^{BC}}$ and
$\zeta = \pm 1$ is the sign of the eigenvalue of \Vtwosq.  The
anticommutator arises on the second line because the holonomy
is a loop, beginning and ending at the vertex where the \Etld
acts. Therefore the grasp occurs on the far left and right of the
holonomy.

    The \Vtwo  in the denominator of the formula \eq{hVPB}
has eigenvectors with vanishing
eigenvalues, and some of these eigenvectors are even
contained in the superposition which forms the coherent state.
Replacing commutator by classical
Poisson bracket will not do for zero eigenvalue states; strictly speaking
one should split
off these states, and calculate their contribution using the original
commutator.

    However, in the classical limit, eigenvectors
of \Vtwo with large values will dominate the coherent state superposition;
contributions from zero eigenvectors
will be small.  Therefore one can split off these eigenvectors--and
ignore their contribution.  \Vtwo is given by its peak value.

    The \Vtwo  factor also does not give rise to a factor ordering
ambiguity, even though it does not commute with the $h^{-1}_x$.
The difference between two orderings equals a commutator, which is
small.
\bea
    h_x \,(1/ \Vtwo) &=& (1/\Vtwo) \,h_x + [ \,h_x, 1/\Vtwo \,];\nn [0 in]
    [ \,h_x , 1/\Vtwo \,] &=& (-1/\Vtwosq) \,[\,h_x,\Vtwo \,] \,h_x \nn [0 in]
                &=& \mbox{order} \,(h_x/\Vtwo)( \,[h_x,\Vtwo]/\Vtwo).
                    \label{hVcommute}
\eea
(On the second line, I assume $h_x$ is outgoing, so that
the \Vtwo operator overlaps with $h_x$ on the left side of
$h_x$.  An ingoing holonomy would give the same final order of
magnitude.)   This commutator
resembles a derivative with respect to L,
in that the commutator lowers the power of L by one.  To see this, note
\Vtwo \hspace{1ex} is order $\sqrt{L_x L_y}$, from the two \Etld operators in \Vtwo.
From \eq{hVPB}, $[ \,h_x,\Vtwo \,]$ is order $L_y/\sqrt{L_x L_y}$, down
by a derivative with respect to $L_x$.  (The $S_B$ is just a Pauli matrix
divided by 2; it is not order L.)  Therefore the two
orderings in \eq{hVcommute} differ by a factor which is suppressed
by  $[ \,h_x,\Vtwo \,]/\Vtwo$ = order $1/L_x$.  In the classical limit,
factor ordering is not a problem.

    I now substitute \eq{hVPB} into the expression for
the trace, \eq{defT}.

\bea
    T(A,x) &=& Tr \,[\,\sigma_A (\sigma_B +h_x \sigma_B h_x^{-1})](1/2)(\gamma\kappa/2) \,
                    \zeta \,\epsilon_{BC} \,\E{y}{C} /\Vtwo\nn
            &=&Tr \,[\,\sigma_A (\sigma_B + \sigma_D D^{(1)}(h_x)_{DB})](1/2)(\gamma\kappa/2) \,
                    \zeta \,\epsilon_{BC} \,\E{y}{C} / \Vtwo \nn
             &=& [ \,\delta_{AB} + D^{(1)}(h_x)_{AB}](\gamma\kappa/2) \,
                    \zeta \,\epsilon_{BC} \,\E{y}{C} / \Vtwo ;\nn
    D^{(1)}(h)&=& D^{(1)}(-\phi+\pi/2,\theta,\phi-\pi/2).
    \label{Vhcomm4}
\eea
The second line uses the rotation property of the su(2) generators,
\eq{Srotn}.  The matrix on the last line is a full-angle rotation, not a
half-angle rotation; it is constructed from $h_x = h^{1/2}_x$, which is in SU(2),
not O(3).

    \Eq{Vhcomm4} is correct for T(A,x). For T(A,y) replace $h_x \,\rta \,
h_y$ and  $\epsilon_{BC} \,\E{y}{C} \,\rta \,\epsilon_{CB} \,\E{x}{C}$.

    A note on commutator = derivative: for the \E{z}{Z} operator,
the classical limit of the commutator is identically the derivative
(with respect to M rather
than L, but a derivative nonetheless).  Compare the commutator, \eq{defTZz},
to the classical limit \eq{limTZz}.  However, for transverse holonomies,
the commutator does not preserve directions.
Compare matrix elements of \Vtwo to those of $ [ \,h^{-1}_x,\Vtwo \,]$,
\eq{hVPB}.  I replace the SU(2) objects $h^{(1/2)}$ by 0(3)
objects $D{(1)}_{0A}$ to facilitate order of magnitude
comparisons:
\begin{eqnarray*}
    <\Vtwo> &\sim& \sqrt{<\E{x}{M}>< \E{y}{N}> \epsilon_{MN}};\\
     <[D{(1)}_{0A},\Vtwo]> &\sim& <D{(1)}_{0B}><\E{y}{N}> \epsilon_{MN}/<\Vtwo>.
\end{eqnarray*}
The net effect is to replace $<\E{x}{M}>$ by $<D{(1)}_{0B}>$.
From \eq{Eeigenval}, $<\E{x}{M}>$
is order $<L_x>$
\[
    <\E{x}{M}> \,= \,<L_x +1/2> (\hat{n}_M \cos \mu -\hat{n}\times \hat{V})_M \sin \mu),
\]
while its replacement is order unity.  Evidently the commutator deletes one
factor of $<L_x>$, in the same manner as a derivative with respect to $<L_x>$.
However, $<\E{x}{M}>$ also
contains a unit vector, and information about this vector has
been lost.  Presumably the commutator = derivative relation was exact for
commutators such as $\sqrt{\E{z}{Z}}$, because only one direction is
involved.  For other operators, the relation is useful when one
is counting powers of L.

\section{Conclusion}\label{SecConclusion}

        The approach used in this paper produces a
clean separation
between leading terms and the small correction terms.
Study of the SC terms, appendix \ref{AppSC},  reveals
that they emphasize small
fluctuations, values of the basic variables M and L
which are near, but not at the mean values $<\mbox{M}>$ and
$<\mbox{L}>$.  The standard deviation parameter
t must be near 1/$<\mbox{L}>$,
or these SC terms will become large.

    The techniques used in
the present, planar calculation may be taken over to
other groups.  The calculation uses only angular momentum
theory, therefore applies
to any symmetry which has recoupling coefficients analogous
to 3J symbols.

    In a follow-on paper the small correction states
are shown to be a complete subset
of the overcomplete set of coherent states \cite{PCSII}.  The SC
states are used to construct a perturbation series
and compute non-leading contributions to the volume operator.

    The next step involves constructing a classical
plane solution, quantizing it, then studying the semiclassical
limit using coherent states.  Since solutions to exact LQG
are hard to construct, it may be necessary to meet the classical
solution halfway: take the semiclassical limit of LQG first;
then quantize the classical theory using
the semiclassical  form.

\appendix
\section{ Identities Involving D and S}\label{Appsphcomp}

    For the convenience of the reader, this appendix includes
brief derivations of two well-known identities involving the
rotation matrices D and generators S.  Throughout, I do not use
the relation $D^{-1} = D\dagger$, so that the formulas in this
appendix remain valid for matrices in SL(2,C).  My conventions
for Clebsch-Gordan coefficients and rotation matrices are
those of Edmonds \cite{Edmonds}

    The first identity relates matrix elements of S to a
Clebsch-Gordan coefficient.

\be
    \bra{L,M'} S_A \,\eta_A \,\ket{ L,M} = \sqrt{L(L+1)}\braket{L,M'}{ L,M;1,A} ,
            \label{SeqCGc}
\ee
where the  Clebsch-Gordan
coefficients for $L1\bigotimes L2 = L3$ are written in a bracket
notation as $\braket{L3,M3}{L1,M1;L2,M2}$.  $\eta_A$ is a phase.
\begin{align}
     n_A &= -1 \, \mbox{for} \, A = +1; \nn
    n_A &= +1 \,\mbox{for} \, A = -1,0.
\label{defeta}
\end{align}
This is the "other" Condon-Shortley phase convention. The operators
$S_A$ are required to have positive matrix elements (the "usual"
convention).  However,
for rotation purposes the $S_A$ form a vector with
spherical components $n_A S_A$.  The indices on Clebsch-Gordan
coefficients have simple rotation properties; therefore
the coefficient in \eq{SeqCGc} is a matrix element of $n_A S_A$.
The next section shows that the factor of $\eta_A$ can be
ignored safely.

The
M dependence of the right hand side of \eq{SeqCGc} is required
by the Wigner-Eckart theorem applied to a spin one
operator.  To check the scalar coefficient $\sqrt{L(L+1)}$, square
both sides and sum over M and A.

\[
    \bra{L,M'} \vec{S}^2 \ket{ L,M'}=L(L+1)\braket{L,M'}{ L,M'}.
 \]
The phase of the scalar coefficient can be verified by checking a
simple example.

    The following identity reduces a product of
two rotation matrices to a single matrix.
(I suppress the labels $L_i$, which are obvious from
context.)

\begin{gather}
    D_{N1 M1} \,D_{N2 M2} = \sum_{L3}\braket{N3}{ N1 \,N2} \,D_{N3 M3} \,
        \braket{M3}{ M1 \,M2}; \nn
    \braket{N3}{ N1 \,N2} \,D_{N1 M1} = D_{N3 M3}
        \braket{M3}{ M1 M2} \,D^{-1}_{M2 N2}.
        \label{DDiden}
\end{gather}
The second line is a rewritten version of the first.

    As an illustration of these rules, I obtain the usual
rotation property for the $S_A$ vector operator.  Insert
\eq{SeqCGc} for S into the second line of \eq{DDiden}, and restore
the L's.

\begin{align}
    \bra{L,N3} S_A \,\eta_A\,\ket{ L,N1} \,D^{(L)}_{N1 M1}
       & = D^{(L)}_{N3 M3} \,\bra{L,M3} S_B \,\eta_B \,\ket{ L,M1} \,D^{(1)-1}_{B A} \nn
       &\Leftrightarrow  \nn
    \mathbf{S}_A \,\mathbf{D}^{(L)} &= \mathbf{D}^{(L)} \,\mathbf{S}_B \,
           \eta_B \, D^{(1)-1}_{B A} \,\eta_A.
                \label{Srotn}
\end{align}
The last line uses a matrix notation to hide some indices.  This formula
is also valid for
matrices D in SL(2,C).

    For components of vectors, this paper uses both Cartesian indices
(X,Y,Z; or 1,2,3) and spherical indices (+1,-1,0).  Strictly speaking, the two should be
treated slightly differently when forming dot products.

\bea
        \vec{A} \cdot \vec{B}&=& A_X B_X + A_Y B_Y + A_Z B_Z \nn
            &=& (A_-)^*  \,B_- + (A_+)^*  \,B_+ + A_3  \,B_3.
                \label{dotproddef}
\eea
Since spherical components are complex, the second dot product resembles
the dot product in
SU(3) rather than O(3).  For the most part I have omitted the complex
conjugation stars, trusting to the reader to know when to put them in.
Note in most (but not all) cases, no star is needed because one of the
indices refers to a final state, which can be considered starred.
For example, there should be a star on one of the factors in
\eq{perpconstraint}; but no star is needed in the sums over magnetic
quantum numbers in \eq{graspcoh1}.

\section{The matrix $\mathbf{D^{(1)}}$(u)}\label{AppD1u}

    This appendix calculates the rows of $D^{(1)}(u)$, which
form a natural basis for the vector operators \Etld and h.
More precisely, from \eq{defC}, the relevant basis vectors
are
\[
        \exp[ \,i \, \beta \, B \,]\, D^{(1)}(u)_{B A}; \quad A,B = +1,0,-1.
\]
The subscripts (B,A) give the components A of unit vector B.

    One may
evaluate the components of each vector, starting from
\[
    D^{(1)}(u)_{B A} = \exp[ \,i(\beta -\pi/2)(A-B) \,] \,d^{(1)}(\alpha/2)_{BA},
\]
with

\be
    d^{(1)}(\alpha/2)= \left[ \begin{array}{ccc}
        (1+\cos)/2& + \sin /\sqrt{2}&(1-\cos)/2\\
        -\sin /\sqrt{2}&\cos & +\sin /\sqrt{2} \\
        (1-\cos)/2& - \sin /\sqrt{2} &(1+\cos)/2
                                     \end{array}\right];
                      \label{d1}
 \ee
$\cos = \cos (\alpha /2), \sin = \sin (\alpha /2)$; rows and
columns are labeled with spherical components in the order
(+1, 0, -1).

    From \eq{DeqHolonomy}, the basis vector B = 0 is just the unit vector
$\hat{D}$. This vector
has spherical and Cartesian components

 \begin{align}
    \hat{D}_A &= D^{(1)}(u)_{0 A} \nn
        &= (\mp \sin(\alpha/2)\exp[\pm i(\beta-\pi/2) \,] /\sqrt{2}, \,
        \cos(\alpha/2))\nn
       &= (\sin(\alpha/2)\sin(\beta), \,-\sin(\alpha/2)\cos(\beta), \,
                \cos(\alpha/2)).
                        \label{Vcomp}
 \end{align}
Spherical components are listed in the order ($\pm$,0); Cartesian
components in the order (X,Y,Z).

    One linear combination
of the rows $B = \pm 1$ turns out to be $\hat{n}$,
the axis of rotation for u, while the
orthogonal linear combination turns out to be
$\hat{n} \times \hat{D}$, the third axis of an orthogonal
coordinate system ($ \hat{n}, \hat{D}, \hat{n}\times \hat{D}$).

\bea
    \sqrt{2}\hat{n}_A &=& -\exp[ \,+i
\beta \,] \,D^{(1)}(u)_{+1, A} +\exp[\,-i \beta \,] \,D^{(1)}(u)_{-1, A}\nn
    &=& -i(d^{(1)}_{+1,A} +
                d^{(1)}_{-1,A})\exp[ \,i A(\beta -\pi/2) \,]\nn
    &=& (\mp \exp[ \,\pm i \beta \,] , \,0) \nn
    &=& \sqrt{2}(\cos(\beta), \,\sin(\beta), \,0);
                \label{ncomp}
\eea
\begin{multline}
    i\sqrt{2}(\hat{n} \times \hat{D})_A = -\exp[ \,+i
\beta \,] \,D^{(1)}(u)_{+1, A} - \exp[\,-i \beta \,] \,D^{(1)}(u)_{-1, A}\\
    = -i(d^{(1)}_{+1,A} -
                d^{(1)}_{-1,A})\exp[ \,i A(\beta-\pi/2) \,]\\
    = i(\mp \cos(\alpha/2)\exp[ \,\pm i(\beta-\pi/2) \,], \,-\sqrt{2}
        \sin(\alpha/2)) \\
    = i\sqrt{2}(\cos(\alpha/2)\cos(\beta-\pi/2),  \,\cos(\alpha/2)\sin(\beta-\pi/2), \,
                -\sin(\alpha/2)).
                \label{nxVcomp}
\end{multline}
For convenience I also record the inverses of
\eqs{ncomp}{nxVcomp}.

\bea
    D^{(1)}(u)_{0 A} &=& \hat{D}(u)_A; \nn
    \exp[ \,+i\beta \,] \,D^{(1)}(u)_{+1, A}&=& (-\hat{n}_A
                    - i (\hat{n}\times \hat{D})_A)/\sqrt{2}; \nn
    \exp[ \,-i\beta \,] \,D^{(1)}(u)_{-1, A}&=& (\hat{n}_A
                    - i (\hat{n}\times \hat{D})_A)/\sqrt{2}.
                    \label{D1=basis}
\eea

    When  $D^{(1)}(u)_{B A}$ is multiplied by the phases
$\eta_B\,\eta_A$, \eq{Srotn}, all factors of  $\mp 1$ in this
section change to +1. (Factors of $\exp (\pm i\,\beta)$ do not change.) 
Therefore \emph{either} include all factors of $\eta$ and drop
the $\mp 1$ factors; \emph{or} ignore factors of $\eta$ and 
retain the factors of $\mp 1$. I have chosen the latter
course in the  body of the paper.

    I could have used the notation $\hat{r}$
for the unit vector which I labeled $\hat{D}$.  The present coherent
states may be used to describe an electron moving in a
central force.  In that interpretation, the
peak value of electron radius  is given by the vector $\hat{D}$.

\section{The matrix $\mathbf{D^{(L)}}$(H)}\label{AppDH}

    This appendix derives the properties of the Hermitean factors
D(H) occurring in the coherent state. The initial formulas
will be valid for general $\hat{p}$ and p; later results will use
the assumptions $e^p \gg 1$ and  $p_3 =  0$.
\[
    \hat{p}= (\cos(\beta + \mu), \sin(\beta + \mu),0).
\]

    Most of the mathematical techniques used in this appendix are
a direct steal from Thiemann-Winkler paper II \cite{GCSII}, with one
significant exception: I make no use of traces.
Because Thiemann and Winkler deal with the
general SU(2) expansion (matrices $D_{MN}$, both M and N summed over)
they are able to recast their results for D(H) as theorems about
class invariants, the traces $D_{MM}$ .  In my case the expansion
matrices are $D_{0M}$, sum over M only; I have not been able to recast my
results as theorems about traces.  Instead, in order to obtain manageable
forms for D(H), I use the assumption $e^p \gg 1$.
This assumption is not a serious limitation
unless one wishes to extend the calculation to very low values of
L of order 10.  For further discussion of this point, see the estimates
given for the size of t, in appendix \ref{AppSC}.  These estimates
in effect also limit the size of p.

    The usual rotation matrices $D^{(L)}$ are finite power series involving
sines and cosines of real angles.  Since these trigonometric functions
are analytic everywhere except at infinity, the real angles can be continued
to complex values, in order to obtain a power series for D(H).  I start from
$D(H)^{(1/2)}$, which has the form

\bea
    D(H)^{(1/2)} &=& \exp [\,\vec{p}\cdot \vec{S}\,] \nn
        &=& \left[ \begin{array}{cc}
        \cosh(p/2) + \hat{p}_3 \sinh(p/2) &
        \sinh(p/2)(\hat{p}_1-i\hat{p}_2)\\
        \sinh(p/2)(\hat{p}_1 +i\hat{p}_2) &
        \cosh(p/2) - \hat{p}_3 \sinh(p/2) \\
         \end{array}\right]\nn
         &:=& \left[ \begin{array}{cc}
         a&b\\
         c&d
        \end{array}\right]
         \label{DHhalf}
\eea
In terms of the abcd, the $D(H)^{(L)}$ for arbitrary L are given
by the finite series

 \begin{multline}
    D^{(L)}(H)_{MN} = \\
                (ad)^L (a/b)^N (b/d)^M \sqrt{(L+M)!(L-M)!(L+N)!(L-N)!}\\
              \cdot  \Sigma_k\frac{(bc/ad)^k}{(L-M-k)!(L+N-k)!(M-N+k)!k!}.
                    \label{Dseries}
\end{multline}

\subsection{Large p limit of D(H)}

    I now assume $e^p$ large.   I am interested in $\hat{p}_3 = 0$,
primarily.  However, in what follows $\hat{p}_3$ can be anything, provided
it is not so close to $\pm1$ that it kills the $e^p$ in the expansions of
a, b, c, and d: $(1-\mid \hat{p}_3 \mid)\gg 2 e^{-p}$.)  Then

\bea
    \cosh(p/2)&\cong& \sinh(p/2)\nn
            &\cong& e^{p/2}/2;\nn
            ad&\cong & bc;\nn
    D^{(L)}(H)_{MN}
                &\cong& (ad)^L(a/b)^N (b/d)^M \sqrt{(L+M)!(L-M)!(L+N)!(L-N)!}\nn
                & & \cdot \Sigma_k
                [(L-M-k)!(L+N-k)!(M-N+k)!k!]^{-1}.
                    \label{Deqseries}
\eea
The series in k can be summed using the addition theorem for
binomial coefficients:
 \be
    \sum_k \left( \begin{array}{c}\mu !\\
                                    k \end{array}\right)
            \left ( \begin{array}{c}\nu ! \\
                                \lambda - k \end{array}\right )
                =
            \left ( \begin{array}{c}(\mu +\nu)! \\
                                    \lambda ! \end{array}\right )
                      \label{binadd}
 \ee
 The \eqs{Deqseries}{binadd} give

\bea
    D^{(L)}(H)_{MN} &\cong& (ad)^L(a/b)^N (b/d)^M \frac {(2L)!}
                        {\sqrt{(L+M)!(L-M)!(L+N)!(L-N)!}}\nn
                    &=&(\hat{p}_1-i\hat{p}_2 )^{M-N} (\exp(p/2)/2)^{2L}
                        [1+\hat{p}_3]^{L+N} [1-\hat{p}_3]^{L-M} \nn
                    & & \cdot \frac {(2L)!}
                        {\sqrt{(L+M)!(L-M)!(L+N)!(L-N)!}}.
                        \label{approxCsh}
\eea

    At this point one can prove: let $\bar{M}$ and $\bar{N}$ denote
the peak values of M and N, i.e. the values which maximize $\mid
D(H) \mid$.  Then

\be
    \bar{M}/L \cong \bar{N}/L \cong \hat{p}_3.
        \label{barMN}
\ee
Proof: to find the peak value of (say) N, compute the first
difference of the square magnitude of the N dependence of D(H),
and set this first difference equal to zero.

\bea
    \delta^{(1)}f(N) &:=& f(N+1) - f(N);\nn
    f(N) &=& \mid a/b \mid ^{2N} /[(L+N)!(L-N)!].\nn
    0 &=& \delta^{(1)}f(\bar{N})\nn
    &\propto&\mid a/b \mid ^2 \frac{L-\bar{N}}{L+\bar{N}+1}
                        -1; \nn
    \bar{N}/L &\cong &\frac{\mid a \mid^2 - \mid b \mid^2}
                                {\mid a \mid^2 + \mid b
                                \mid^2};\nn
    \bar{N}/L &\cong & \hat{p}_3.
        \label{barNcalc}
\eea
On the last line I have used the values of abcd from \eq{DHhalf}.
The proof  for $\bar{M}$ is identical except for the replacements
(a,b) $\rta \,$(b,d). $\Box$

\subsection{Small $\mathbf{p_3}$ limit of D(H)}

    In the main body of the text I focus on the case
$p_3 = 0$.  From \eq{barMN} of the last subsection, in this limit
the important values of M and N satisfy $L>> M,N $.  Therefore one
can use Stirling's approximation for the factorials in
\eq{approxCsh}, for example

\be
    \frac {(2L)!}
        {(L+M)!(L-M)!} \cong \frac {(2L^{2L}}
        {\sqrt{\pi}(L-M)^{L-M+1/2}(L+M)^{L+M+1/2}}.
\ee
Now use

\bea
    (1+x/n)^n &= & \exp[n \ln(1+x/n)] \nn
            &\cong& \exp [x - x^2/2n + \ldots].
            \label{powereqexp}
\eea
Take $n = L + 1/2$, $x = \pm M$.  Also, write $L^{2L}$ as
$(L+1/2 -1/2)^{2L}$ and apply \eq{powereqexp} to this factor.
 \be
\frac {(2L)!}
        {(L+M)!(L-M)!} \cong \frac {(2^{2L} \exp[-M^2/(L+1/2) \,]}
                                {\sqrt{\pi (L+1/2)}}
                                \label{GaussianM}
\ee
To obtain a result valid near $p_3 = 0$, as well as at
$p_3 = 0$, assume $\hat{p}_3 \leq
\mbox{order} \,1/\sqrt{L+1/2}$.  Then apply  \eq{powereqexp}
to the $[1\pm \hat{p}_3]$ factors, with now n = L+1/2, x = $\pm
\hat{p}_3 \,(L+1/2)$ . For example,

\begin{align}
    [1 + \hat{p}_3]^{L+N} &=
        [1 + \hat{p}_3 (L+1/2)/(L+1/2)]^{(L+1/2)[1+(N-1/2)/(L+1/2)]}\nn
    &\cong \exp[\, \hat{p}_3(L+1/2)+ \hat{p}_3(N-1/2)  \nn
       & \quad - (\hat{p}_3)^2(L+1/2)/2 \,],
    \label{p3eqexp}
\end{align}
and similarly for the $[1 -
\hat{p}_3]$ factor.  Inserting
\eqs{GaussianM}{p3eqexp} into \eq{approxCsh} yields

\bea
    D(H)^{(L)}_{MN}& \cong &(\hat{p}_1-i\hat{p}_2 )^{M-N} \frac{\exp(p L)}{\sqrt{\pi (L+1/2)}}\nn
            & & \cdot \exp\{-[M-\hat{p}_3(L+1/2)]^2 /2(L+1/2)\}\nn
            & & \cdot \exp\{-[N-\hat{p}_3(L+1/2)]^2 /2(L+1/2)\}
                \label{approxD}
\eea
The M (and N) dependence of D(H)
is peaked at $M = \hat{p}_3(L+1/2)$, with the squared width of the
Gaussian equal to $\sqrt{L(L+1)} \cong (L+1/2)$.  This is already
a bit more than we need for the main body of the paper.

    \Eq{approxD} demonstrates Gaussian behavior in N and M.  To
obtain Gaussian behavior in L, multiply D(H) by the other
exponential factor in the coherent state.

\begin{multline}
    \exp(-t L(L+1)/2 \,) \,D(H)^{(L)}_{NM} \cong \exp(-t
    L(L+1)/2 \,) \,e^{Lp}\cdots  \\
        = \exp[-t((L+1/2)-p/t)^2 /2 +
                p^2/(2t) +t/8 -p/2 \,]\cdots .
                \label{GaussianL}
\end{multline}
The $\cdots$ indicates irrelevant factors which are bounded for
large L. On the last line one can neglect $\exp(t/8)\cong 1$.
\Eq{GaussianL} is a Gaussian in L with mean $<L+1/2> = p/t$ and
standard deviation $1/\sqrt{t}$.

    Since t is small, the standard deviation is very large.  However,
what counts is (standard deviation)/(mean value of variable),
\[
    \sigma _L /<L+1/2> = \sqrt{t}/p ,
\]
which is small as required.

    When $p_3 = 0$, it is parameterized by the
angle $\mu$ introduced at \eq{defmu}.  $\mu$ is the
angle between $\hat{n}$ and $\hat{p}$.
\[
    \hat{p}= (\cos (\beta + \mu), \,\sin(\beta + \mu), \,0)
\]
For this value of $\hat{p}$, D(H) becomes

\begin{gather}
    \exp(-t L(L+1)/2) \,D(H)^{(L)}_{MN} \cong
            \exp[-t((L+1/2)-p/t)^2 /2 \,]
             \nn
     \cdot \exp [-M^2 /2(L+1/2) \,] \exp [-N^2 /2(L+1/2) \,]
                [1/\sqrt{\pi (L+1/2)}]\nn
     \cdot \exp [ \,p^2/(2t) -p/2 \,](\exp[ \,-i(\beta + \mu) \,])^{M-N}.
            \label{DHforp3eq0}
\end{gather}

 \section{Small correction (SC) terms}\label{AppSC}

This appendix discusses the nature of the small corrections SC.
\[
    \mbox{operator}\ket{\mbox{coh state}} =  \mbox{$<$operator$>$}\ket{\mbox{coh
    state}}
             + SC.
\]
The coefficients multiplying these states
are shown to be suppressed
by factors involving the small parameters $1/\sqrt{<L>}$ and $\sqrt{t}$.

    Appendix \ref{Appt}  argues that the parameter
t should be taken to be order $1/<L>$, in order to minimize
the size of the SC terms.  If t is replaced by a number of order
$1/<L>$, everywhere in the factors multiplying the SC terms,
then all the SC terms turn out to be suppressed by factors of the same
order, $1/\sqrt{<L>}$.

    The SC terms emphasize values of the
parameters L and M which are near, but not at the average value.
I.~e. the dynamical variables of the theory connect the
original coherent state not  only to itself, but also to coherent states with
peak values near those of the original coherent state.

\subsection{SC states for the {\boldmath $\tilde{E}$ }operator}

    At \eq{graspcoh3a} I replaced
\[
    L \,\rta \,<L +1/2> + \Delta L; \,M \,\rta \,\Delta M,
\]
then asserted that the terms proportional to $<L +1/2>$ represented the
dominant contribution.  I must now examine the terms involving
$\Delta X$, X = L or M, and show that they are small.

    From the
previous appendix, \eq{DHforp3eq0}, the original coherent state
is proportional to Gaussian factors coming from the D(H) factor.
Therefore the SC terms are proportional to the first moments
of Gaussians.
\begin{gather}
    \ket{u, \vec{p}}\propto \sum_{L,M} D^{(L)}(h u\dagger)_{0M}\nn
               \cdot \exp[-t((L+1/2)-p/t)^2 /2 \,] \exp[-M^2 /2(L+1/2)\,];\nn
     \mbox{SC terms} \propto \sum_{L,M} D^{(L)}(h u\dagger)_{0M} \,[ \,\Delta L \:\mbox{or}\: M \,]\nn
                 \cdot \exp[-t((L+1/2)-p/t)^2 /2 \,] \exp[-M^2 /2(L+1/2) \,].
\label{ArgumentOfSC}
\end{gather}

    The  original coherent state has the standard Gaussian
form, with single peaks at $\Delta L = 0 \:\mbox{and}\: M$ = 0.
The SC states have a zero where
the original state has a peak; and a peak plus valley at two points
located a standard deviation away from the original peak.

    The SC states resemble the difference between two Gaussians, each peaked
at a value near, but not at the original peak.  As mentioned
earlier, the operators connect the coherent state to
itself, but also to nearby coherent states.

    It is difficult to carry out the sums over L and M in
\eq{ArgumentOfSC} because the
$D^{(L)}(h u\dagger)_{0M}$ factor is
difficult to approximate.
There is a simpler way to estimate the order of magnitude of the
SC terms, without knowing in detail the M and L dependence of
$D^{(L)}(h u\dagger)_{0M}$. For the SC terms involving M and
$\Delta L$, define the states

\bea
    \ket{1M} &:=& N(1(M))\sum_{L,M}
            ((2L+1)/4\pi) \exp[-t L(L+1)/2 \,]\nn
           & & \cdot [ \,D^{(L)}(h u\dagger)_{0M} \,M \,D^{(L)}(H) \,]_{M0};
                \label{defm1M}
\eea
\bea
    \ket{1L } &:=& N(1( L))\sum_{L,M}
            [(2L+1)/4\pi] \exp[-t L(L+1)/2 \,]\nn
           & & \cdot [ \,D^{(L)}(h u\dagger)_{0M}\,\Delta L \,D^{(L)}(H) \,]_{M0}.
                \label{defm1L}
\eea
The notation pX denotes the pth moment of the variable X.  The above
states are identical to the original coherent state $\ket{u, \vec{p}}$
except for a different normalization factor,
\[
    N \,\rta \,N(1X),
\]
and one power of $\Delta X$ in the summand.  In terms of these states,
\eq{graspcoh3a} becomes

\bea
    (\gamma\kappa/2)^{-1}\E{x}{A}\ket{u,\vec{p}} &=&
        <L +1/2> \,(\hat{n}_A \cos \mu -\hat{n}\times \hat{D})_A \,\sin \mu) \ket{u,\vec{p}}\nn
      & &  + \,(N/N(1L))\ket{1( L)} \nn
    & &    + \,(N/N(1M)) \,[ \,\hat{D}_A +i \sin \mu \,\hat{n}_A \nn
      & &  + \,i\cos \mu \,(\hat{n}\times \hat{D})_A \,]\ket{1M} .
            \label{graspcoh4}
\eea
Evidently this replaces the problem of evaluating the sums over L,M
by the problem of determining the normalization ratios N/N(1X).
This may seem like replacing Tweedledum by Tweedledee, except
the dangerous factors of $D^{(L)}(h u\dagger)_{0M}$
drop out when calculating norms.

        I now prove the following:
\begin{gather}
     1 \cong  N^2 \exp[ \,p^2/t -p \,] \,\sqrt{<L+1/2>}/2\pi \sqrt{t};\nn
     N/N(1M) \cong \sqrt{<L+1/2>/2}: \nn
     N/N(1L) \cong \sqrt{1/(2 \,t)}.
            \label{NoverNi}
\end{gather}

I begin with the first line of \eq{NoverNi}.  Orthogonality
for the D(h) is
\[
    \int_{\Omega(h)} \,D^{(L)}_{0M}(h) \,D^{(L')}_{0M'}(h)^* =
            \delta_{L,L'} \,\delta_{M,M'} \,(2L+1)/4 \pi.
\]
From this and \eqs{defcoh}{geqHu},

\bea
    1&=& \braket{u,\vec{p}}{u,\vec{p}}\nn
     &=& N^2\sum_{L,M}\exp[-t L (L+1) \,] \,[(2 \,L+1)/4 \pi \,] \,
                    D^{(L)}_{M0}(g\dag)^* \,D^{(L)}_{M0}(g\dag) \nn
        &=& N^2
        \sum_{L,M,N,N'}\exp[-t L(L+1) \,] \,[(2 \,L+1)/4\pi \,] \nn
         & &  \cdot     D^{(L)}_{0N}(H) \,D^{(L)}_{NM}(u) \,D^{(L)}_{MN'}(u\dag) \,D^{(L)}_{N'0}(H)  \nn
        &=& N^2\sum_{L,M}[(2 \,L+1)/4\pi \,] \exp[-t L(L+1) \,] \nn
        & & \cdot D^{(L)}_{0M}(H) \,D^{(L)}_{M0}(H).
        \label{Ncalc1}
\eea
As advertised earlier, the difficult u and h dependence has disappeared.

    The next step is to carry out the sum over M.
Compare \eqs{defh}{geqHu}: when going from
$h^{1/2}$ to $H^{(L)}$ we make the replacements
\[
    i \,\hat{m} \cdot \vec{S} \,\theta \,\rta \,\vec{S}\cdot \vec{p}.
\]
I.e. replace magnitude and direction as follows.
\[
    i \,\theta \,\rta \,p; \quad \phi \,\rta \,\beta + \mu
        \qquad (\Leftrightarrow i \,\vec{m} \,\rta \,\vec{p}).
\]
The angles are defined at \eqs{defm}{defmu}.
Therefore the Euler decomposition of D(H) follows from the
Euler decomposition of D(h), \eq{defh}.
\begin{multline*}
      D(h) \,(-\phi +\pi /2, \theta ,\phi -\pi/2)  \\
           \rta \,D(H) \,(-\beta -\mu+\pi/2,-ip,\beta +\mu -\pi/2).
\end{multline*}
Therefore the D's on the last line of \eq{Ncalc1} equal
\[
    D^{(L)}(-\beta -\mu +\pi/2, \,-2ip, \,\beta +\mu -\pi/2)_{00}.
\]
Approximate this factor using
\eq{DHforp3eq0} with $p \rta 2p, t \rta 2t$.

\be
    1 \cong N^2 \sum_L [(2L+1)/4\pi]\frac{\exp[-t((L+1/2)-p/t)^2  +
                p^2/(t) -p \,]}{\sqrt{\pi (L+1/2)}}.
            \label{Ncalc2}
\ee
Replace the sum over L by an integral:

\bea
    \sum_L (\Delta L = 1) &= &(1/\sqrt{t})\sum_L \Delta
    (\sqrt{t}(L+1/2)-p/\sqrt{t}:=w) \nn
        &\cong &(1/\sqrt{t})\int dw.
            \label{sumLeqint}
\eea
Elsewhere in the integral, replace L + 1/2 by its peak value
\[
    <L+1/2> = p/t.
\]
\Eq{Ncalc2} then gives the first line of \eq{NoverNi}.

    The remaining two lines of \eq{NoverNi} may be proved using
similar approximations, with one exception.   The calculation of
N(1M) resembles the calculation of N, \eq{Ncalc2}, except for
an additional factor of $M^2$ in the summand, so that
the sum over M cannot be carried out immediately.
Instead, the sum over M may be replaced by an integral over
a variable q, using

\bea
    \Sigma_M (\Delta M = 1) & =& \sqrt{L+1/2}\,\Sigma (\Delta
    M/\sqrt{L+1/2}:= \Delta q) \nn
                &\cong &\sqrt{L+1/2}\int dq.
                \label{DMeqw}
\eea
$\Box$

    \Eq{NoverNi} implies that the SC terms are suppressed: from
\eq{graspcoh4}, the leading
term is order $<L>$; therefore the SC terms are down by factors of order
\begin{eqnarray*}
    N/[ \,N(1M)<L>\,] &=&  1/\sqrt{2<L>}; \\
    N/[ \,N(1L)<L> \,] &=& 1/(\sqrt{2t}<L>).
\end{eqnarray*}

At first glance the formula for N/N(1L) looks dangerous, because of
the small factor of $\sqrt{t}$ in the denominator.  However,
$t = p/<L+1/2>$, from \eq{DHforp3eq0}.  Therefore
\[
        N/N(1L)/<L>  \,\cong  \,1/\sqrt{2 p <L>}.
\]
Since I am taking
$e^p \gg 1$, both SC terms are down by at least $1/\sqrt{<L>}$.

\subsection{SC states for the holonomy}

    As with the \Etld operators, I estimate the order of magnitude of
the SC terms by calculating appropriate norms.  For the \Etld
operators, the SC terms were proportional to new states $\ket{1X}$
which resemble the original coherent state, except for an additional
factor of $\Delta X$.   Since the holonomy
produces states containing $D^{(L')}(h)$ with $L'=L \pm 1$, the
states  $\ket{1X}$ are not enough, and I will need the following
additional states:

\bea
    \ket{L_{\pm}} &:=& N(L_{\pm})\sum_{L,M}((2L+1)/4\pi) \exp[-t
                                    L(L+1)/2 \,]\nn
       & &\cdot (1/2) \,[ \,D^{(L+1)}(hu\dagger)\pm D^{(L-1)}(hu\dagger) \,]_{0M} \,
                        D^{(L)}(H)_{M0};\nn
       \ket{L+1,1M}&:=& N(L+1,1M)\sum_{L,M}((2L+1)/4\pi) \exp[-t
                                    L(L+1)/2 \,]\nn
       & &\cdot D^{(L+1)}(hu\dagger)_{0M} \,M \,
                        D^{(L)}(H)_{M0}.
        \label{defLpm}
\eea
In terms of these states, \eq{hcoh3} becomes

\bea
    D^{(1)}(h)_{0,A}\ket{u,\vec{p}} &\cong& N/N(L_+)\ket{L_+}\hat{D}_A \nn
            & &+ N/N(L+1,1M)\ket{L+1,1M} \nn
            & &\cdot [\,-\cos \mu \,\hat{n}_A
                            +\sin \mu \, (\hat{n}\times \hat{D})_A \,]/<2L+1)> \nn
           & & -i N/N(L_-)\ket{L_-} [ \,\sin \mu \,\hat{n}_A + \cos \mu \,(\hat{n}\times \hat{D})_A \,]\nn
        & &+   \quad \mbox{order} \,(M/L)^2,1/L.
                \label{hcoh4}
\eea
The ratios which determine the order of magnitude
of the SC terms are

\bea
    N/N(L_+) &\cong& 1;\nn
    N/N(L_-) &\cong& \sqrt{t/2};\nn
    N/N(L+1,1M) &\cong& \sqrt{<L+1/2>/2} .
        \label{NoverNiforh}
\eea
Proof: the last line of \eq{NoverNiforh} is easiest to establish.  Because
$\ket{\mbox{L+1,1M}}$ differs from $\ket{1M}$ only in the
replacement of $D^{(L+1)}(hu\dagger)$ by $D^{(L)}(hu\dagger)$, and the
$D(hu\dagger)$ factors drop out anyway when computing norms, N/N(L+1,m1M)
is the same as N/N(m1M), \eq{NoverNi}.

    To determine the N($L_{\pm}$), rewrite the $L_{\pm}$
states as follows.

\bea
    \ket{L_{\pm}} &=& N(L_{\pm})\sum_{L,M}((2L+1)/4\pi) \exp[-t
                                    L(L+1)/2 \,]\nn
       & &\cdot (1/2)[ \,D^{(L+1)}(hu\dagger)\pm D^{(L-1)}(hu\dagger) \,]_{0M} \,
                        D^{(L)}(H)_{M0}\nn
       &\cong& N(L_{\pm})\sum_{L,M}((2L+1)/4\pi)D^{(L)}(hu\dagger)_{0M}\nn
        & &\cdot (1/2)\sum_{\pm}[(\pm 1)\exp\{-t[L \mp 1+1/2 -p/t \,]^2/2 \,\}]\nn
         & & \cdot \exp\{ \,p^2/2t-p/2-i M (\beta+ \mu) \,\}\frac{\exp\{-M^2/[2(L +1/2) \,]\}}{\sqrt{\pi(L
        +1/2)}}\nn
        &\cong& N(L_{\pm})\sum_{L,M}((2L+1)/4\pi) \,D^{(L)}(hu\dagger)_{0M}\nn
        & &\cdot \exp\{-t[L +1/2 -p/t \,]^2/2 \,\}\left[\begin{array}{c}
                \cosh [ \,t(L+1/2-p/t)]\\
                \sinh [ \,t(L+1/2-p/t)]
                     \end{array}
            \right]\nn
         & & \cdot \exp\{ \,p^2/2t-p/2-i M (\beta+ \mu) \,\}\frac{\exp\{-M^2/[2(L +1/2) \, ]\}}{\sqrt{\pi(L
        +1/2)}}\nn
        &\cong& N(L_{\pm})\sum_{L,M}((2L+1)/4\pi) \,D^{(L)}(hu\dagger)_{0M}\nn
         & & \cdot \exp[-t L(L+1)/2 \,] \,D^{(L)}(H)_{M0} \nn
                                          & &  \cdot    \left[\begin{array}{c}
                                                \cosh [ \,t(L+1/2-p/t)]\\
                                                \sinh [ \,t(L+1/2-p/t)]
                                                 \end{array}\right].
            \label{LpmCalc1}
\eea
The cosh (sinh) goes with the upper (lower) sign.
On the third line I have relabeled $L \pm 1 = \tilde{L}$, used \eq{DHforp3eq0} to
replace the D(H) by Gaussians, and then dropped the tildes.
From the last line, the $\ket{L_{\pm}}$ states are just the
original states times an additional factor of cosh or sinh.  When
the state is squared to determine a norm, this factor becomes
$\cosh^2 [\sqrt{t}w]$ or $\sinh^2 [\sqrt{t}w]$, as at
\eq{sumLeqint}.  The variable L is replaced by a variable w, and
the sum over L is replaced by an integral over w . The
w dependence of the Gaussian gives w $\leq$ 1, and t is
order 1/$<L>$.  For the $L_+$ state,
$\cosh^2[\sqrt{t}w]
\cong 1$.  The normalization integral for N($L_+$) reduces to
the normalization integral for N, and we get the first line of
\eq{NoverNiforh}.  For the $\mbox{L}_-$ state, $\sinh^2 [\sqrt{t}w]\cong t w^2$.
This normalization integral should be compared to the normalization integral for
$\ket{1L}$.  That integral contains a $(\Delta L)^2 = w^2/t$ factor.
Therefore in  $(N/N(1L))^2=  1/(2 t)$, \eq{NoverNi}, move the t from
denominator to numerator to get  $(N/N(L_-))^2 =  t/2$. $\Box$

\section{Estimates of the parameter t} \label{Appt}

    The parameter t, introduced at \eq{defcoh}, is analogous to the
standard deviation parameter $\sigma$ present in the coherent state
for the free particle, \eq{freecoh}.  That parameter drops out
of the Heisenberg relation $\Delta p \Delta x \geq \hbar$, but
does determine the individual uncertainties $\Delta x \cong \sigma$,
$\Delta p \cong \hbar/\sigma$.  One can choose extreme values of
$\sigma$ leading to "squeezed"  states.

    It is difficult to put significant limits on the parameter t, if
one looks only at leading terms.  The coefficients of the leading
terms are the peak values, and the only peak value
(of holonomy, \Etld, L, or angles) which depends on t is $<L>$.
From \eq{DHforp3eq0} even
this peak value depends on t only via p/t, rather than p alone.
\be
    <L+1/2> \,= \,p/t.
    \label{avgL}
\ee
    To put limits on t, one must consider the SC terms.
I list  various states contributing to
the SC terms.  First, the \Etld SC terms,
from \eqs{graspcoh4}{NoverNi}:
\[
    \sqrt{1/(2<L>)}\, \ket{1M}; \sqrt{1/(2t<L>^2)} \,\ket{1L}.
\]
Each state is smaller than the leading term by the factor
multiplying the state.
Next, the holonomy SC terms, from \eqs{hcoh4}{NoverNiforh}:
\[
    \sqrt{1/4<L>} \,\ket{L+1,1M}; \sqrt{t/2} \,\ket{L_-}
\]
One of these terms has t in the numerator, and one has t in the
denominator.  we can determine
a best value of p and t by summing these two factors
\[
    \sqrt{1/(2t <L>^2)}+\sqrt{t/2},
\]
then minimizing the sum with respect to t. The resulting best value is
\be
    t = 1/<L>.
    \label{bestt}
\ee

    I cannot set t = 1/$<L>$ exactly, however, because then from \eq{avgL} I
must take p = 1.  Appendix \ref{AppDH} requires an expansion in
$\exp (-p) \ll 1$ to neglect non-leading  terms in D(H).  As a compromise, I take p
large, but not too large; say p = 5.  Then the expansion of D(H)
remains valid, since $\exp(-5)$ is small; also the SC terms
will be small, provided
$<L+1/2>$ is large enough.  The t dependent factors suppressing the SC terms
become
\[
    \sqrt{1/(2t<L>^2)}=\sqrt{1/p<L>}; \, \sqrt{t/2}=\sqrt{p/2<L>}.
\]
For $<L> $ greater than 100 or so, and p = 5, these factors are
sufficiently small.

    When the SC terms are taken into
account, t is not arbitrarily adjustable.  Values of p and t much
different from p = 1 and t = 1/$<L>$ result in
larger-than-optimal SC terms.

\end{document}